\documentclass[aps,prb,superscriptaddress,twocolumn,floatfix,citeautoscript]{revtex4-1}

\usepackage{amsmath}
\usepackage{amssymb}
\usepackage{amsfonts}
\usepackage{dsfont}
\usepackage{graphicx,graphics}
\usepackage{color}
\usepackage[autostyle]{csquotes}

\begin{document}

\title{ Injection locking and synchronization in Josephson photonics devices }
\author{Lukas Danner}
\email{lukas.danner@uni-ulm.de}
\affiliation{Institute of Quantum Technologies, German Aerospace Center (DLR), S\"oflinger Stra{\ss}e 100, D-89077, Ulm, Germany}
\affiliation{Institute for Complex Quantum Systems and IQST, Ulm University, Albert-Einstein-Allee 11, D-89069 Ulm, Germany}
\author{Ciprian Padurariu}
\email{ciprian.padurariu@uni-ulm.de}
\affiliation{Institute for Complex Quantum Systems and IQST, Ulm University, Albert-Einstein-Allee 11, D-89069 Ulm, Germany}
\author{Joachim Ankerhold}
\email{joachim.ankerhold@uni-ulm.de}
\affiliation{Institute for Complex Quantum Systems and IQST, Ulm University, Albert-Einstein-Allee 11, D-89069 Ulm, Germany}
\author{Bj\"orn Kubala}
\email{bjoern.kubala@uni-ulm.de}
\affiliation{Institute of Quantum Technologies, German Aerospace Center (DLR), S\"oflinger Stra{\ss}e 100, D-89077, Ulm, Germany}
\affiliation{Institute for Complex Quantum Systems and IQST, Ulm University, Albert-Einstein-Allee 11, D-89069 Ulm, Germany}
\date{\today}

\begin{abstract}

Injection locking can stabilize a source of radiation, leading to an efficient suppression of noise-induced spectral broadening and therefore, to a narrow spectrum. 
The technique is well established in laser physics, where a phenomenological description due to Adler is usually sufficient.
Recently, locking experiments were performed in Josephson photonics devices, where microwave radiation is created by inelastic Cooper pair tunneling across a dc-biased Josephson junction connected in-series with a microwave resonator. An in-depth theory of locking for such devices, accounting for the Josephson non-linearity and the specific engineered environments, is lacking.

Here, we study injection locking in a typical Josephson photonics device where the environment consists of a single mode cavity, operated in the classical regime. We show that an in-series resistance, however small, is an important ingredient in describing self-sustained Josephson oscillations and enables the locking region. We derive a dynamical equation describing locking, similar to an Adler equation, from the specific circuit equations. The effect of noise on the locked Josephson phase is described in terms of phase slips in a modified washboard potential. For weak noise, the spectral broadening is reduced exponentially with the injection signal. When this signal is provided from a second Josephson device, the two devices synchronize. In the linearized limit, we recover the Kuramoto model of synchronized oscillators. The picture of classical phase slips established here suggests a natural extension towards a theory of locking in the quantum regime. 

\end{abstract}

\maketitle

\section{Introduction}\label{sec:intro}


A number of experiments have demonstrated injection locking in solid state devices that generate microwave radiation, 
e.g. double quantum dot masers \cite{Liu2015} and Josephson microwave amplifiers \cite{Huard2019}. 
Recently, injection locking has also been observed in a Josephson photonics device dubbed the Josephson laser \cite{Cassidy2017} refining earlier experiments employing a Cooper pair transistor \cite{Chen2014}. 
Josephson photonics devices use inelastic Cooper pair tunneling across a Josephson junction
to convert the energy from a constant voltage bias source into microwave radiation of corresponding frequency.
By applying a small amplitude ac signal on top of
the dc voltage bias, a small `locked' region was found in Ref.~[\onlinecite{Cassidy2017}], 
where the spectrum of the emitted microwave radiation becomes an extremely sharp peak pinned to  
the frequency of the injected signal.
Injection locking may prove a key enabling technology for Josephson photonics devices and, as we will show here, also allows to study interesting synchronization phenomena when realized by coupling different devices.

The typical setup of Josephson photonics devices
consists of a voltage-biased Josephson junction coupled in-series
 to a microwave cavity. The cavity acts 
as an  environment engineered to absorb the energy of the tunneling Cooper pair as microwave photons \cite{Hofheinz2011}.
These 
devices have been developed as efficient sources of \emph{quantum} microwaves by tailoring
the properties of the cavity, or by deploying several cavities in series. 
In that manner, recent experiments have demonstrated 
bright single photon sources \cite{Grimm2019, Rolland2019} as well as
sources of two-mode light with nonclassical correlations \cite{Westig2017} and of entangled photons \cite{Peugeot2020}.
Theoretically, Josephson photonics devices have been proposed as platforms for non-linear dynamics experiments
that take advantage of the strong Josephson non-linearity and can be operated both in the classical and in the 
quantum regimes \cite{Padurariu2012,Gramich2013,Armour2013,Juha2013,Kubala2015,Armour2015,Trif2015,Meister2015,Souquet2016,Juha2016,
Armour2017,Wang2017,Simon2018,Juha2018,Arndt2019,Morley2019,Kubala2020,Lang2020} by designing low- or high-impedance resonators.

The scope of these varied applications can be significantly broadened by enhancing the phase stability of the emitted light against electronic noises, currently an outstanding challenge. 
While phase stability is an ubiquitous issue in microwave photonics of all types, Josephson photonics devices are particularly susceptible because the radiation is generated using a constant voltage source. This does not provide a reference phase and thus  favors no particular value of the oscillation phase, i.e., in the language of non-linear dynamics 
the phase is neutral. 

Here, we will show that injection locking is an efficient method to achieve phase stability of Josephson photonics circuits.
In this approach, an ac signal is 
injected directly into the circuit,
with the purpose of phase locking the coupled Josephson and cavity oscillations.
Furthermore, injection locking 
straightforwardly extends to a scenario, where an ac-signal stemming from one Josephson photonics circuit is used to lock another one, so that the two devices synchronize.
While exploring injection locking and synchronization as paradigmatic examples of nonlinear classical dynamics in a new system class is of interest in itself, Josephson photonics devices will allow future studies to extend such investigations to a regime
where the dominant source of noise is due to quantum fluctuations. In particular, as we will later argue, injection locking in these devices could
become a promising platform for studying 
quantum tunneling of the Josephson phase, a phenomenon analogous to 
 the flux tunneling, or quantum phase slips \cite{Hriscu2013}, observed in thin superconducting wires \cite{Bezryadin2000,Altomare2006,Astafiev2012,Chen2014b}. 

The textbook theoretical description of 
injection locking is based on 
a phenomenological approach due to Adler \cite{Adler1964}. 
 Applied to a universal class of oscillators, termed self sustained oscillators, the Adler equation describes their effective dynamics by a single equation for the (limit cycle) phase in the locked region and its vicinity \cite{Pikovsky2001}.
Experimentally observed features in Ref.~[\onlinecite{Cassidy2017}], such as  frequency pulling and a scaling of the locking region's width with the
amplitude of 
the injection signal, can successfully be described by such an approach.
However, a derivation of the effective Adler equation starting from the fundamental underlying equations-of-motion, the Josephson circuit equations, has not been established. In consequence,  the crucial parameter of the Adler model, the critical detuning that defines the locked region, has not been connected to the device properties which in turn means that the optimal design of Josephson photonics devices for locking remains unknown.

In fact, the Josephson laser in Ref.~[\onlinecite{Cassidy2017}] 
operated in a large dc-voltage bias regime, where microwaves are emitted at a down-converted frequency from a multi-mode cavity, and the resulting dynamics of the device is complicated and highly non-linear. Recently, a theoretical description of the device dynamics was formulated \cite{Simon2018}, that describes the microwave emission using an approach based on a semiclassical system of coupled non-linear differential equations. However, the model used could not capture the observed injection locking phenomena. The reason for this could not be clearly identified and to the best of our knowledge 
has remained an open challenge until now. The experiment Ref.~[\onlinecite{Cassidy2017}] and the theory of Ref.~[\onlinecite{Simon2018}] have been taken to imply that the highly non-linear microwave emission generated in the Josephson laser and 
possibly the presence of a multi-mode cavity, may be a requirement for observing injection locking. 
As we show here, this interpretation turns out to be inaccurate,
emphasizing the need for theoretical understanding of injection locking in Josephson photonics devices, particularly the theoretical description of the simplest Josephson photonics device that can be injection locked. Absent such a description,
the best design, the circuit parameters, and the bounds on noises required to observe injection locking have not been systematically identified for Josephson photonics devices.

\begin{figure}[t]
\begin{center}
\includegraphics[width=0.99\columnwidth]{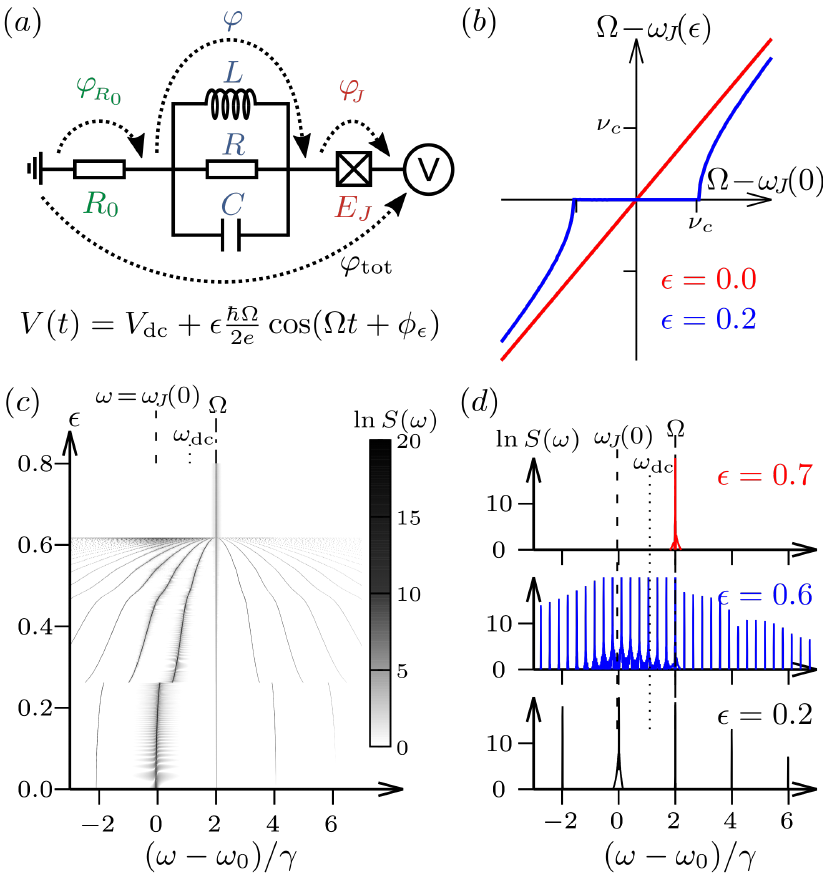}
\caption{
(Color online.) Injection locking of Josephson oscillations. (a) Sketch of a Josephson photonics circuit with an in-series resistance and a single resonance (with frequency $\omega_0=1/\sqrt{LC}$ and width $\gamma=1/RC$).
(b) Typical Adler curve showing injection locking [$\omega_J(\epsilon)=\Omega$] as the Josephson frequency $\omega_J(\epsilon)$ is pulled towards the locking frequency $\Omega$.
Without a locking signal the Josephson frequency $\omega_J(0)$ is determined by the voltage bias, $\omega_\textrm{dc}=2eV/\hbar$, reduced by the voltage drop at the in-series resistance $R_0$ [$\omega_J(0)\approx \omega_0+2 \gamma$; $\omega_\textrm{dc} =\omega_0+2.3\gamma$].
Locking occurs for detunings, $\Omega - \omega_J(0)$, below a critical value $\nu_c(\epsilon)= \frac{e}{\hbar} I_c R_0 \epsilon$, see Sec.~\ref{sec:injection}.\\
(c) Grayscale plot of emission spectrum $S(\omega,\epsilon)$ and
(d) cuts for various locking signal amplitudes $\epsilon$. 
In the locked state all emission occurs at the locking frequency $\Omega$, whereas without a locking signal the cavity emits at the Josephson frequency given by the (reduced) dc-bias, $\omega_J(0)$. The transition is marked by a characteristic fan structure. [Parameters values for $\omega_\textrm{dc},\omega_J(0)$  in (c) and (d) differ from (b) as indicated on the figure;  other parameters: $Q=\omega_0/\gamma=30$, $\frac{2e}{\hbar} I_c R/\omega_0 =  \tilde{I}_c = 0.5 $, $\frac{2e}{\hbar} I_c R_0 = v_{R_0}= \frac{4}{30} \omega_0$.]
}
\label{fig:intro}
\end{center}
\end{figure}

Here, we study injection locking and synchronization in Josephson photonics devices in the classical regime. 
We show that injection locking can be achieved in the simplest Josephson photonics device, 
where the environment consists of an LC-resonance (a single mode cavity) and a (small) in-series resistance $R_0$, see Fig.~\ref{fig:intro}(a).
We mainly focus on the case when the bias voltage 
is set close to this resonance, $2eV_\textrm{dc}/\hbar = \omega_\textrm{dc}  \approx \omega_0=1/\sqrt{LC}$, and only briefly discuss the situation where
the radiation is a result of parametric down conversion, when the bias is tuned near twice the value of the resonance frequency. 
For this model, we derive analytically the device-specific Adler-type equation starting
from the Kirchhoff equations of the circuit. Based on this Adler equation, we provide bounds on the electric noises that can be stabilized, 
and suggestions for optimal device design.

In Sec.~\ref{sec:pre} we will analyze the semi-classical non-linear behavior of our model circuit based on the Kirchhoff equations and
explain the concepts of forced versus self-sustained oscillations. In Sec.~\ref{sec:injection} we derive the analytical 
equation of Adler-type that describes the injection locking region and the slow dynamics of the Josephson phase just outside the locked region. 
The effects of noise, the appearance of phase slips and the reduction of the line width are presented in Sec.~\ref{sec:noise}.  
 Section~\ref{sec:higher_order} deals with locking at higher-order resonances, 
while in Sec.~\ref{sec:synch} we turn to the synchronization of two capacitively coupled copies of the Josephson photonics device. 
Finally, Sec.~\ref{sec:concl} presents our concluding remarks and an outlook on future work.

\section{Preliminaries}\label{sec:pre}


We study a Josephson photonics device modeled by the circuit in Fig.~\ref{fig:intro}(a). The dynamics of the circuit is conveniently
parametrized by the phase drop over the Josephson junction $\varphi_J$ and the phase across the LC-resonator $\varphi$.
The classical equations of motion are the Kirchhoff equations of the circuit
\begin{subequations}
\begin{align}
\dot\varphi_J = \frac{2e}{\hbar}V(t) - \frac{2e}{\hbar}I_c R_0 \sin(\varphi_J) - \dot{\varphi};\label{eq:eom1}\\
\ddot\varphi+\gamma\dot\varphi+\omega_0^2\varphi = \frac{2e I_c R}{\hbar\omega_0} \gamma\omega_0 \sin {\varphi_J}.\label{eq:eom2}
\end{align}
\label{eq:eom}
\end{subequations}
Here, $\omega_0 = 1/\sqrt{LC}$ is the resonance frequency and $\gamma= 1/RC$ is the resonance width. The parametrization of physical quantities characterizing the device used frequently throughout the paper is organized in Table~\ref{table:1}.

\begin{figure}[t]
\begin{center}
\includegraphics[width=0.99\columnwidth]{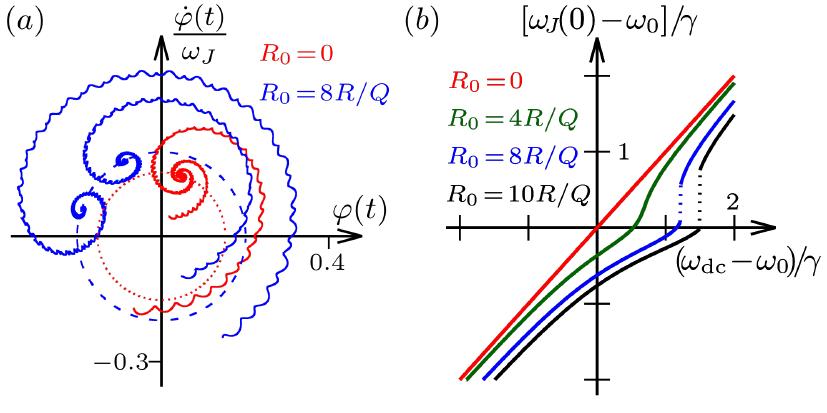}
\caption{(Color online.) Properties of Josephson oscillations without a locking signal. (a) Phase space
picture (in the frame rotating with $\omega_J$) of the cavity dynamics showing trajectories (solid) relaxing towards limit cycles (dashed). Without in-series resistance a Josephson photonics circuit is forced into oscillations with a fixed, stable phase, so that all perturbations relax back to the same phase space (fixed) point (red, dots: lab-frame). The additional degree of freedom provided by an in-series resistance allows different perturbations to relax to different points along the limit cycle (blue). This freedom of the phase on the limit cycle makes a self-sustained oscillator susceptible to locking. Note that, in the reduced phase space of the cavity, trajectories may cross.
(b) The oscillation frequency of the cavity, $\omega_J(0)$, is reduced from the dc-bias, $\sim \omega_\textrm{dc}$, by the (dc-)voltage drop at the in-series resistance. For small $R_0$, this reduction is given by the Lorentzian resonance curve of Josephson photonics (red vs. green curve), whereas at larger $R_0$ an instability occurs.\\ 
 {[}Parameters: $\omega_\textrm{dc} = \omega_0 + 1.5 \gamma$, $0\le \omega_0 t \le 450$ in (a); for other parameters see Fig.~\ref{fig:intro} and legends.]
}
\label{fig:prelim}
\end{center}
\end{figure}

Before showing in Sec.~\ref{sec:injection} in detail how Eq.~\eqref{eq:eom} allows the phase of the radiation emitted by the circuit
 to be stabilized by a locking signal, we want to gain some physical intuition, how the Josephson nonlinearity impacts the dynamics described by Eq.~\eqref{eq:eom} within or outside-of the locked regime. 
 
While nonlinearity is generically known as a crucial ingredient for locking and synchronization phenomena, nonlinear effects have also been extensively studied (theoretically and experimentally) for Josephson photonics circuits, 
where $R_0 \equiv 0$ and locking phenomena are absent.
In that case Eq.~\eqref{eq:eom1} can be directly integrated (setting $\varphi_J(0) + \varphi(0)=0$) and substituted in \eqref{eq:eom2} resulting in 
$\ddot\varphi+\gamma\dot\varphi+\omega_0^2\varphi = \tilde{I}_c \gamma\omega_0 \sin{\left[ \frac{2e}{\hbar}\int_0^t dt' V(t') - \varphi \right]}$, the standard equation-of-motion of  (classical) Josephson-photonics (cf. Refs.~[\onlinecite{Armour2013}]~and~[\onlinecite{Meister2015}]) describing a \emph{harmonic} LC-resonator with an \emph{unconventional nonlinear drive term}. 

Considering a pure dc-drive close to the fundamental resonance, $\omega_\text{dc}- \omega_0 \lesssim \gamma $, it is easy to see, that for small $\tilde{I}_c \ll 1$, the oscillation amplitude of $\varphi(t)$ remains small and the zeroth-order expansion of the Josephson-nonlinearity in  $\varphi$ yields a linearly driven harmonic oscillator, that responds with
an oscillation described by a complex amplitude $\tilde{\varphi}$ with $|\tilde{\varphi}| = \left| \tilde{I}_c \gamma/[ 2(\omega_\text{dc} - \omega_0) +i \gamma]\right|$. 

For larger $\tilde{I}_c$ a rotating-wave approximation yields a time-independent equation for the oscillation amplitude at frequency $\omega_\text{dc}$ (higher harmonics of $\omega_\text{dc}$ are off-resonant and can be neglected)
\begin{align*}
    \tilde{\varphi} = \tilde{I}_c \frac{\gamma}{ 2(\omega_\text{dc} - \omega_0) +i \gamma}\left(J_0(|\tilde{\varphi}|)\frac{\tilde{\varphi}^*}{|\tilde{\varphi}|}-J_2(|\tilde{\varphi}|)\frac{\tilde{\varphi}}{|\tilde{\varphi}|}\right).
\end{align*}
The Bessel functions of the first kind are manifestations of the highly non-linear Josephson drive and can give rise to bifurcations and multi-stable states in this regime.

\begin{table}[t]
\centering
\begin{tabular}{l @{\hskip 0.3cm} c @{\hskip 0.3cm} l} 
 \hline
 Physical Quantity & Parametrization & Unit \\ [0.5ex] 
 \hline
 resonance frequency & $\omega_0$ & s$^{-1}$ \\
 resonance width    & $\gamma=(RC)^{-1}$ & s$^{-1}$\\
 d.c. bias voltage & $\omega_\text{dc}=(2e/\hbar) V_\text{dc}$ & s$^{-1}$ \\
 Josephson frequency & $\omega_J\neq \omega_\text{dc}$ & s$^{-1}$ \\
 injection signal frequency & $\Omega$ & s$^{-1}$ \\
 scale of residual voltage at $R_0$ & $v_{R_0}=(2e/\hbar) I_c R_0$ & s$^{-1}$ \\
 Josephson driving strength & $\tilde{I}_c = (2e/\hbar) I_c R/\omega_0$ & 1 \\
 injection signal amplitude & $\epsilon=(2e/\hbar)V_\text{ac}/\Omega$ & 1 \\
 \hline
\end{tabular}
\caption{Physical quantities that characterize the device.}
\label{table:1}
\end{table}

Adding a small ac-signal to the voltage drive,
\begin{align}
V(t)= V_\text{dc} + V_\text{ac} \cos(\Omega t+\phi_\epsilon),
\label{eq:Vac}
\end{align}
 leads to a simple superposition of responses with the two frequencies $\omega_\text{dc}$ and $\Omega$, if $R_0$ is neglected.

In most of the following, to emphasize the simplest locking scheme, we restrict our study to locking of the linear regime of Josephson-photonics, where  $\tilde{I}_c \ll 1$.
Despite the small amplitude of resonant oscillations in this regime, it is the presence of the in-series resistance $R_0$ which will lead to a different manifestation of Josephson nonlinearity and pave the path to locking and synchronization phenomena. To understand the nature of this distinct nonlinear effect, we have to consider how the presence of a finite resistance $R_0$ modifies the picture of the driven oscillator sketched above.

\paragraph*{Origin of locking:} For finite but small $R_0 \ll R$, and $\tilde{I}_c\ll 1$  so that $\varphi$ remains small, Eq.~\eqref{eq:eom1} can be iteratively integrated 
\begin{flalign}
&\varphi_J(t) = \int^t_0 \!\! dt' \left\{\frac{2e}{\hbar}V(t')  - v_{R_0} \sin{[\varphi_J(t')]}\right\} - \varphi(t) \label{BORN-like}\\ 
&\approx \frac{2e}{\hbar}\int_0^t  \!\! dt' V(t')  - \varphi(t) 
-v_{R_0}\!\!  \int_0^t \!\! dt_1  \sin\left[\frac{2e}{\hbar} \! \int\nolimits_0^{t_1} \!\! dt_2 V(t_2)\right] \nonumber
\end{flalign}
in a Born-like approximation up to leading order in the small quantities $\tilde{I}_c$ and $(v_{R_0}/\omega_0)$. [Note also that $\varphi$ is small with $|\tilde{\varphi}|\propto \tilde{I}_c$.] 

In addition to the dc- and ac-components of the voltage drive, the presence of $R_0$ gives rise to a low-frequency Fourier component $\tilde{\varphi}_J(\nu) = {\cal{F}}\!\left[ \varphi_J(t)  \right]_{\nu}$, at the frequency of beats $\nu = (\omega_\text{dc} - \Omega)$ caused by a slowly oscillating part of the Josephson current $I_J \sim \sin\varphi_J$. 
In fact, we find
\begin{align}
\left| \tilde{\varphi}_J(\nu) \right|=& v_{R_0}\,  {\cal{F}}\!\left[ \int_0^t dt_1  \sin\left(\frac{2e}{\hbar} \int\nolimits_0^{t_1} dt_2 V(t_2)\right)
 \right]_{\nu} \nonumber\\
 =& \frac{\epsilon}{2}\frac{v_{R_0}}{\nu}\,,\nonumber
\end{align}
 i.e., \emph{the closer the frequencies of locking signal $\Omega$ and dc-drive $\omega_\text{dc}$ become, the slower the oscillations in $I_J$, and the larger the integrated effect in the  $\tilde{\varphi}_J(\nu)$ oscillations.}
 
In consequence, even for small locking amplitude $\epsilon$ and small, but finite, $v_{R_0}$ the low-frequency oscillations in the Josephson phase  $\tilde{\varphi}_J(\nu)$ 
can become so large, that (despite $\tilde{I}_c\ll 1$) the nonlinearity of the driving term $\sim\sin\varphi_J$ in Eq.~\eqref{eq:eom2}
comes to bear. This nonlinearity of the slow-response produces an increasing number of sidebands in the oscillator response $\tilde \varphi$ around the Josephson frequency, each sideband separated by $\nu$, as seen in Fig.~\ref{fig:intro}(d), until eventually the oscillations lock, with the response concentrated at frequency $\Omega$, cf. the typical fan-structure of Fig.~\ref{fig:intro}(c). 

While in Sec.~\ref{sec:injection}, we will properly derive the corresponding locked and unlocked solution, from the simple arguments above one can already estimate the locking range; namely by the onset of the new type of nonlinearity as $1 \overset{!}{=}  \left|\tilde{\varphi}_J(\nu) \right| = (\epsilon/2) (v_{R_0}/\nu)$ which yields a locking region 
\begin{align}
|\nu| = | \Omega - \omega_\textrm{dc} | \leq \nu_c =\frac{1}{2} \epsilon v_{R_0}. \label{eq:locking_region}
\end{align}
Considering the crucial role any residual in-series resistance $R_0$ plays in establishing the nonlinear locking dynamics, it may seem surprising that it was mostly neglected in previous studies of Josephson-photonics.
This is justified by the observation that without an ac-signal  the residual resistance merely leads to a small and constant shift of the dc voltage with no significant consequence for the dynamics \footnote{It has been noted, however, that a model without $R_0$ and a strictly fixed dc-bias would result in emission without spectral width, and the observed spectral linewidth, typically much sharper than the inverse cavity lifetime, has been associated with low-frequency fluctuations of the voltage at the junction}.
Only in combination with an additional injection signal, the residual resistance becomes a key parameter enabling low-frequency oscillations to access the nonlinear regime. The shift of the dc voltage, $I_\text{dc} R_0$, becomes dependent on this nonlinear response $\varphi(t)$, so that the effective dc-voltage seen by the junction and the resulting Josephson frequency is pulled and the Josephson oscillations can lock to the ac-signal. 
\paragraph*{Forced and self-sustained oscillations:} 
In the language of non-linear dynamics, without $R_0$ the Josephson junction undergoes \textit{forced oscillations}, where the Josephson frequency is directly fixed to the dc bias. The forced oscillations respond linearly to an injection signal. At finite $R_0$, the Josephson frequency becomes a dynamical quantity, different from the dc bias, that is determined by the equation 
\begin{align}
\label{eq:Jos_freq}
\omega_J = \omega_\text{dc} - \frac{2e}{\hbar} I_\text{dc}[ \varphi\,; \varphi_J ] R_0,
\end{align}
 where the dc part of the Josephson current,  $I_\text{dc}$, depends on the solutions $\left\{ \varphi(t), \varphi_J(t)\right\}$. 
In this case, the Josephson oscillations are \textit{self-sustained oscillations} and can be locked.

These properties of a Josephson photonics device with an in-series resistance are demonstrated in Fig.~\ref{fig:prelim}. 
The cavity degree of freedom $\varphi$ is driven to oscillations with $\omega_J < \omega_\textrm{dc}$ according to \eqref{eq:Jos_freq}. In the weak driving regime, $\tilde{I}_c \ll 1$, these are purely sinusoidal and correspond to a single point in Fig.~\ref{fig:prelim}(a) in a frame rotating with $\omega_J$. Depending on initial conditions different angles on a circular limit cycle are found in the long-time limit. Perturbations from this (non-equilibrium) steady state relax to a (possibly different) point on the limit cycle. These are all the typical features of a self-sustained oscillator, such as the van-der-Pol oscillator, with the slightly more complex relaxation dynamics here being caused by the extra degree of freedom $\varphi_J$. In contrast, without $R_0$, oscillations occur with   $\omega_J \equiv \omega_\textrm{dc}$ and a fixed phase angle (only depending on the choice of the rotating frame \footnote{Note, that there is a subtlety in defining the phase of the rotating frame and there is also a valid approach which results in a limit cycle description for the $R_0=0$ case, but does not lead to locking.
}) that corresponds to a single point on a cycle, stable against perturbations. 

Figure~\ref{fig:prelim}(b) visualizes how the in-series resistance $R_0$ impacts the frequency of Josephson oscillations even in the absence of a locking signal. Without $R_0$ the Josephson frequency is given by the dc bias ($\omega_J = \omega_\text{dc}$, red), whereas as described by Eq.~\eqref{eq:Jos_freq} for finite $R_0$ it is reduced by the (average) voltage drop at the in-series resistance $R_0$ due to the dc-part of the Josephson current $I_\text{dc}$. For small $R_0$ (green) this reduction is simply given by the standard resonance curve of Josephson photonics, where $I_\text{dc}$ (and the photon emission) peaks at the fundamental cavity resonance $\omega_\text{dc} = \omega_0$ with a resonance width determined by the cavity decay rate $\gamma$. Increasing $R_0$ (blue, black) the resonance is shifted to larger $\omega_\text{dc}$ compensating for the portion of the voltage dropping at $R_0$. For voltages above the resonance peak, i.e., at the flank of the resonance curve where $Y_J=(2e/\hbar) \partial I_\text{dc}/ \partial \omega_J < 0$, an instability develops, once $Y_J \le -1/R_0$, so that the total device admittance vanishes. Formally, this follows by differentiating Eq.~\eqref{eq:Jos_freq}  with respect to $\omega_J$, where $\partial \omega_\text{dc}/ \partial \omega_J = 0$ corresponds to the jumps in Fig.~\ref{fig:prelim}(b).


\section{Injection locking}\label{sec:injection}


In this Section, we derive explicit equations describing the non-linear behavior developing at small detuning $\nu \simeq \nu_c\ll\omega_J,\Omega$ and show that it results in frequency pulling and locking. In particular, we derive the equations governing the dynamics of the circuit at the slow frequency $\nu$ and show the similarity to the Adler equation, a universal equation describing injection locking in self-sustained oscillators. 

Slow dynamics will describe how the circuit response shifts from oscillations at the natural frequency of the self-sustained oscillator $\omega_J$ to the frequency $\Omega$ of the locking signal. 
The definition of $\omega_J$ by Eq.~\eqref{eq:Jos_freq} applies in the presence of a locking signal, so that $\omega_J=\omega_J(\epsilon)$ with $\omega_J(\epsilon) \equiv \Omega$ when the oscillations are locked.

In the limit of weak Josephson coupling $\tilde{I}_c \ll 1$ and small residual resistance $R_0\ll R$, 
we can neglect the components of $\varphi_J$ oscillating at 
multiples of the Josephson frequency. However, for a small frequency $\nu$ when phase locking develops, we must account for sidebands around $\omega_J$ at multiples of $\nu$. We express these sidebands by  slowly-varying parameters in an ansatz for junction and cavity phases.
Since locking primarily relies on the adaption of frequencies and phases, we can justifiably neglect a slow time dependence of oscillation amplitudes and write
\begin{align}
\varphi_{J}(t) = \omega_J t + \theta_J(t) + a_J \sin[\omega_J t+\phi_J(t)].\label{eq:phiJs}
\end{align}
to capture the driven dynamics of the junction phase consistent with our considerations above, cf.~Eq.~(\ref{BORN-like}). For the cavity phase we similarly chose the ansatz,
\begin{align}
\varphi(t) = a \sin[\omega_J t+\phi(t)]\; ,\label{eq:phis}
\end{align}
which neglects the off-resonant response of the cavity phase, assuming a large quality factor, $Q\gg 1$, 
as is typical in Josephson photonics devices.

\paragraph*{Time scale separation:} 
All unknown functions, $\theta_J(t),\,\phi_J(t)$ and $\phi(t)$, are slowly-varying in time, with $\dot\theta_J,\,\dot\phi_J,\,\dot\phi\simeq\nu\ll\omega_J$, while the `fast' 
frequencies are only slightly detuned from each other, $\omega_J\simeq\omega_\textrm{dc}\simeq\omega_0$.
We can thus use time scale separation for the two timescales $\omega_J^{-1}\ll \nu^{-1}$
to obtain equations for the slowly varying functions. 
Inserting the ansatz in the circuit equations Eqs.~\eqref{eq:eom1} and \eqref{eq:eom2}, we can separate each resulting equation in a slow part and a part containing fast oscillations. The relation
\begin{align}
\dot\theta_J =&\ \left(\omega_\text{dc} - \omega_J\right) - v_{R_0}\frac{a_J}{2}\sin(\phi_J-\theta_J) ,\label{eq:slow1}
\end{align}
follows from the low frequency terms of Eq.~\eqref{eq:eom1}, while the corresponding result from Eq.~\eqref{eq:eom2} yields an irrelevant off-set of the $\phi$ oscillations.

Isolating terms with frequencies close to $\omega_J$ in Eq.~\eqref{eq:eom1} and Eq.~\eqref{eq:eom2}, respectively, yields
\begin{widetext}
\begin{subequations}
\begin{align}
\omega_Ja_J\cos(\omega_Jt+\phi_J) =&\ \epsilon\Omega\cos(\Omega t+\phi_\epsilon)-v_{R_0}\sin(\omega_Jt+\theta_J)- a\omega_J\cos(\omega_Jt+\phi) ,\label{eq:slow2}\\
\tilde{I}_c\omega_0\gamma \sin(\omega_Jt+\theta_J) =&\ a\;\left\{ \left[\omega_0^2-\left(\omega_J+\dot\phi\right)^2\right]\sin(\omega_Jt+\phi)+
\gamma\omega_J\cos(\omega_Jt+\phi)\right\}.\label{eq:slow3}
\end{align}
\end{subequations}
\end{widetext}

Equations \eqref{eq:slow2} and \eqref{eq:slow3} can be rewritten in a frame rotating with frequency $\omega_J$, where they are equivalent to the following
set of complex equations,
\begin{subequations}
\begin{align}
a_J e^{i\phi_J} =&\ \epsilon e^{i\phi_\epsilon}e^{i(\Omega-\omega_J)t} + i\frac{v_{R_0}}{\Omega}e^{i\theta_J} - a e^{i\phi} ,\label{eq:slow2complex}\\
a e^{i\phi} =&\ \tilde{I}_c\frac{\gamma}{\left[2\left(\omega_0-\omega_J-\dot\phi\right)+i\gamma\right]}e^{i\theta_J}.\label{eq:slow3complex}
\end{align}
\end{subequations}
Eqs.~\eqref{eq:slow2complex} and \eqref{eq:slow3complex}, together with Eq.~\eqref{eq:slow1}, form a set of five real-valued algebraic equations
that can be solved for three unknown real-valued slowly-varying functions $\theta_J$, $\phi_J$, and $\phi$, and two real-valued
parameters $a_J$ and $a$.

\paragraph*{Deriving an Adler equation:}  
The injection locking phenomenon emerges by solving the above system of equations.
However, the full analytical solution is cumbersome and fortunately, its behavior can be discussed 
quantitatively using a simplification commonly used in non-linear sciences. 
The simplification arises when the system of equations is reduced to a single Adler-type equation,
where a locking region can be directly identified. 
To derive the Adler-type equation analytically,  we proceed 
by substituting Eq.~\eqref{eq:slow3complex} into Eq.~\eqref{eq:slow2complex},
then taking the imaginary part of Eq.~\eqref{eq:slow2complex} to obtain an expression for $a_J\sin(\phi_J-\theta_J)$. 
Finally, the latter expression
can be substituted into Eq.~\eqref{eq:slow1} to arrive at
\begin{widetext}
\begin{align}
\dot\theta_J = \left(\omega_\text{dc} - \omega_J\right) - \frac{v_{R_0}}{2}\left[\epsilon \sin\left[(\Omega-\omega_J)t-\theta_J+\phi_\epsilon\right]+
\frac{v_{R_0}}{\Omega}+\tilde{I}_c\frac{\gamma^2}{4\left(\omega_0-\omega_J-\dot\phi\right)^2+\gamma^2}\right] .\label{eq:Adler1}
\end{align}
\end{widetext}
Eq.~\eqref{eq:Adler1} captures the dynamics of the circuit at the slow timescale $\nu^{-1}$, describing the non-linear 
response of the circuit to the injection signal. The analogy to the Adler
equation becomes evident by defining the phase $\psi(t)$ as 
\begin{align}
\psi(t) = \omega_Jt+\theta_J - \Omega t-\phi_\epsilon.\label{eq:AdlerPsi}
\end{align}
In the phase locked regime, where $\theta_J=\text{const.}$ and $\omega_J=\Omega$, the phase $\psi(t)$ 
becomes constant in time, $\dot\psi=0$. 
Substituting the expression of $\psi$ into Eq.~\eqref{eq:Adler1} we arrive at 
\begin{align}
\dot\psi =&\ \nu(\epsilon,\Omega)  -v_{R_0}\;\frac{ \epsilon}{2}\sin(\psi),\label{eq:Adfin}
\end{align}
with the detuning $\nu(\epsilon,\Omega)$ defined as 
\begin{align}
\nu(\epsilon,\Omega) \equiv \nu_0(\Omega)+\Delta\nu(\epsilon,\Omega),
\end{align}
where $\nu_0$ is the detuning between the injection frequency $\Omega$ and the Josephson frequency at $\epsilon=0$, 
i.e. before the injection signal is applied, and $\Delta\nu(\epsilon,\Omega)$ is the self-consistent change in the Josephson frequency, owing to the 
frequency-dependence of the cavity response. These two terms have the following expressions,
\begin{widetext}
\begin{align}
\nu_0(\Omega) =&\ \left(\omega_\text{dc} -  \frac{v_{R_0}}{\Omega}\: \frac{v_{R_0}}{2} - \frac{\tilde{I}_c}{2} \frac{\gamma^2}{4\left[\omega_0-\omega_{J}(0)\right]^2+\gamma^2}\: v_{R_0} \right)- \Omega. \\
\Delta\nu(\epsilon,\Omega) =&\ \frac{\tilde{I}_c}{2}\left[\frac{\gamma^2}{4\left[\omega_0-\omega_{J}(0)\right]^2+\gamma^2}-
\frac{\gamma^2}{4\left[\omega_0-\omega_J(\epsilon,\Omega)-\dot\phi(\epsilon,\Omega)\right]^2+\gamma^2}\right]\: v_{R_0}.\label{eq:deltanu}
\end{align}
\end{widetext}

The detuning $\nu$ that depends on both the injection amplitude $\epsilon$ as well as frequency $\Omega$,
together with the prefactor of the sine term $(\epsilon/2) v_{R_0}$, reflect an Adler-type equation 
that is specific to Josephson photonics circuits. This is a central result of this paper.

\begin{figure}[t]
\begin{center}
\includegraphics[width=0.99\columnwidth]{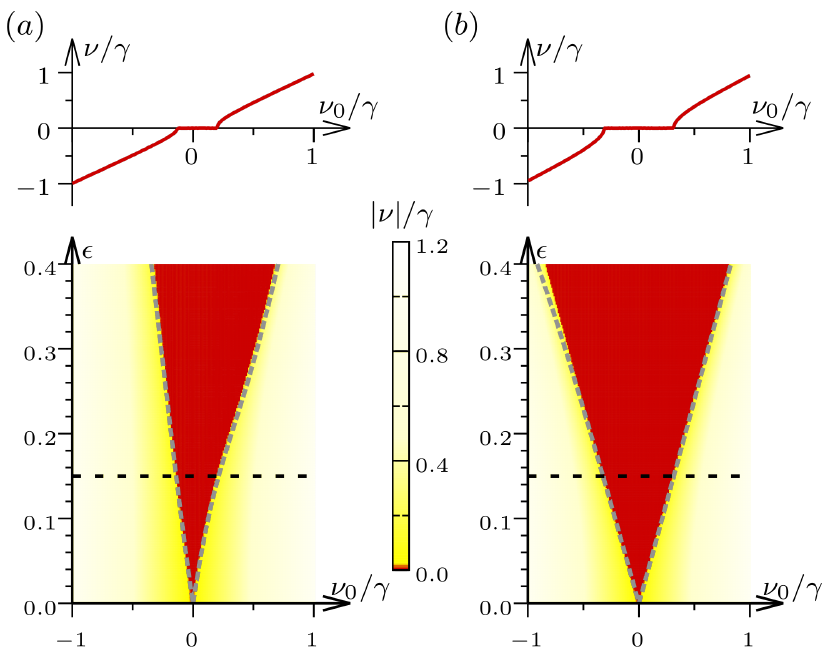}
\caption{(Color online.) Arnold tongues. 
In the locked region (red area in lower panels and $\nu \equiv 0$ in upper panels) the Josephson frequency, $\omega_J(\epsilon)$ becomes equal to the locking frequency, $\nu=\omega_J - \Omega \equiv 0$. The initial ($\epsilon=0$) detuning, $\nu_0 =  \omega_J(0) - \Omega$, is varied by changing the locking frequency $\Omega$ for a constant dc-bias.\\ (a) If the initial Josephson frequency is close to the cavity resonance [$\omega_{J}(0) \approx \omega_0 -  \gamma/20$ found for $\omega_\textrm{dc} = \omega_0 + 0.39 \gamma$] the Arnold tongue is strongly deformed for $\epsilon \not\ll 1$, while (b) for off-resonant biasing [$\omega_{J}(0) \approx \omega_0 +2 \gamma$ found for $\omega_\textrm{dc} = \omega_0 + 2.3 \gamma$] the standard linear Arnold tongue shape is recovered. Black dashed lines indicate the position of the horizontal cuts shown in the upper panels. Grey dashes are the analytical results for the locking boundaries based on the derived effective Adler-equation, Eqs.~\eqref{eq:Adfin}-\eqref{eq:deltanu}. [The Josephson driving strength is (a) $\tilde{I}_c=0.3$ and (b) $\tilde{I}_c=0.5$. Other parameters are the same as in Fig.~\ref{fig:intro}.]
}
\label{fig:tongues}
\end{center}
\end{figure}

\paragraph*{Arnold tongues as locked regions:}  
The Adler equation, Eq.~\eqref{eq:Adfin}, reveals the phase-locked region marked by solutions where $\dot\psi(t)=0$. Such solutions
exist for any fixed injection signal amplitude $\epsilon$, as long as the injection frequency $\Omega$ is tuned such that 
$|\nu(\epsilon,\Omega)|<(\epsilon/2) v_{R_0}$. In that case, we find the time-independent solution $\psi$ given by,
\begin{align}
\psi = \arcsin\left(\frac{2\nu(\epsilon,\Omega)}{\epsilon v_{R_0}}\right) \quad (\textrm{mod }2\pi).
\end{align}
The edges of the phase-locked region are determined by the self-consistent equation,
\begin{align}
|\nu(\epsilon,\Omega)|= \frac{1}{2} \epsilon v_{R_0}. 
\end{align}
The resulting phase-locked region typically increases as a function of $\epsilon$, creating a shape termed the Arnold tongue. In Fig.~\ref{fig:tongues}
we show such Arnold tongues, first in the case when the dc bias is chosen such that the Josephson frequency in the absence of injection signal is near
the resonance of the cavity, $\omega_J(0)\approx \omega_0$, and second, in the case when the frequencies are far from the cavity resonance frequency.

Away from the resonance, in the frequency range where the electromagnetic environment of the junction is feature-less, the term $\Delta\nu(\epsilon)$ is suppressed and can
be neglected if $R\ll R_0$. In this limit, the cavity can be neglected and the circuit becomes equivalent to a resistively  shunted Josephson junction (RSJ). 
For the RSJ model, Eq.~\eqref{eq:Adfin} takes the canonical form of the Adler equation,  
\begin{align}
\dot\psi =&\ \nu_0 - v_{R_0}\;\frac{ \epsilon}{2}\sin(\psi).\label{eq:Shapiro}
\end{align}
For a sufficiently strong drive and small detuning $\nu_0$, such that $\nu_0\leq \nu_c$ with $\nu_c\equiv v_{R_0}( \epsilon/2)$,
the Adler equation admits a time-independent solution $\psi_0=\arcsin(\nu/\nu_c)$ 
that describes locked oscillations, $\dot\varphi_J = \Omega$, with relative phase $\psi_0$. 
For the RSJ model, the phase locked region describes the well known Shapiro step \cite{Shapiro1963},
occurring at  $2e\langle V_J \rangle = \hbar\Omega$ between current plateaus  in the I(V) curve.

Alternatively to an oscillatory voltage, injection locking can also be achieved by injecting an ac-current directly into the cavity of the Josephson device. The roperties of locking are qualitatively similar to those described in this section. For completeness and due to the possible relevance for future experimental implementations, we have described in detail the derivation of the effective Adler equation for direct ac-current locking in the Supplementary Material (S.M.).

\paragraph*{Washboard potentials and locking signatures:}  
It may be helpful to emphasize at this point the difference between two washboard-like potentials for the dynamics of different phase-like variables which appear in the locking scenario.
First, the washboard potential for the dynamics of the Josephson phase $\varphi_J$, obtained by Eq.~\eqref{eq:eom1}, describes a potential with a periodic part given by the Josephson coupling, and a tilt which oscillates in time with an amplitude proportional to $\epsilon$ around an average value set by the dc-voltage (with an additional term describing the coupling to the cavity variable). Without ac-drive and cavity, this is, of course, just the familiar RSJ-potential for an overdamped junction describing, e.g., transitions between a running-voltage state and the trapped state. Second, the Adler-like potential, Eq.~\eqref{eq:Adfin}, also of washboard type, associated to the dynamics of the Adler-variable $\psi$. This potential can essentially be related to the one for $\varphi_J$ by a rotating-frame description (with rotation frequency $\Omega$), where the tilt becomes static and is reduced to the detuning, and the periodic part stems from a combination of locking signal and Josephson coupling. 

While the details of Eqs.~\eqref{eq:Adfin}-\eqref{eq:deltanu} and the way locking affects the cavity emission are unique to the Josephson-photonics setup, many aspects of the dynamics in the tilted washboard potential are generic: 
The dynamics of a particle with negligible mass in a washboard potential $U(\psi) = \nu\psi + \nu_c\cos(\psi)$ depends on the relation between tilt and oscillation amplitude. The condition $\nu \leq \nu_c$ corresponds to the appearance of local minima in the potential that trap the relative phase $\psi$ and stabilize it. 
Just outside the locking region, $\nu \gtrsim \nu_c$, 
$U(\psi)$ has the shape of a staircase. The phase particle
advances slowly along the flat region of the staircase, then drops rapidly along the steep region, giving rise
to periodic dynamics (for $\psi$ modulo $2\pi$). 
We define the
distance from the locking region, $\delta = (\nu-\nu_c)/\nu_c$ and assume $\delta\ll 1$. 
The time spent by the phase particle on the flat region of the staircase is given by $T_s \sim (2\pi/\nu_c) (1/\sqrt{\delta})$ and
diverges as the locking region is approached, $\delta\rightarrow 0$.
The rapid drop along the steep region is relatively short, $(2\pi/\nu_c)$, and can be neglected in comparison to $T_s$
close to the locking region.
The slow rise during $T_s$ followed by a rapid drop is well described by a sawtooth function. Its period is given by the slow timescale $T_s$
and can be represented as a Fourier series in harmonics of $\tilde\nu \simeq (2\pi/T_s) \sim \nu_c\sqrt{\delta}$,
\begin{align}
\left[\psi(t) - \psi(0)\right]_{(\text{mod}\; 2\pi)} \sim \sqrt{\delta} \sum_{k=1}^{1/\sqrt{\delta}} (-1)^{k+1} \frac{\sin\left(k \tilde\nu t\right)}{k},
\label{eq:sawtooth}
\end{align}
valid close to the locking region, in the asymptotic limit $\delta\rightarrow 0$.
The emergence of many, closely spaced harmonics $k \tilde\nu$ with $\tilde\nu  \sim \sqrt{\delta}$ and $k=1 \hdots 1/\sqrt{\delta}$ for  $\delta\rightarrow 0$ in the dynamics of the rotating-frame variable $\psi$ results in the fan-like spectrum of the cavity emission centered around $\omega_J$ observable in Fig.~\ref{fig:intro}(c),(d), a telltale signature of the approach to the locking transition.

\begin{figure*}[t]
\begin{center}
\includegraphics[width=0.99\textwidth]{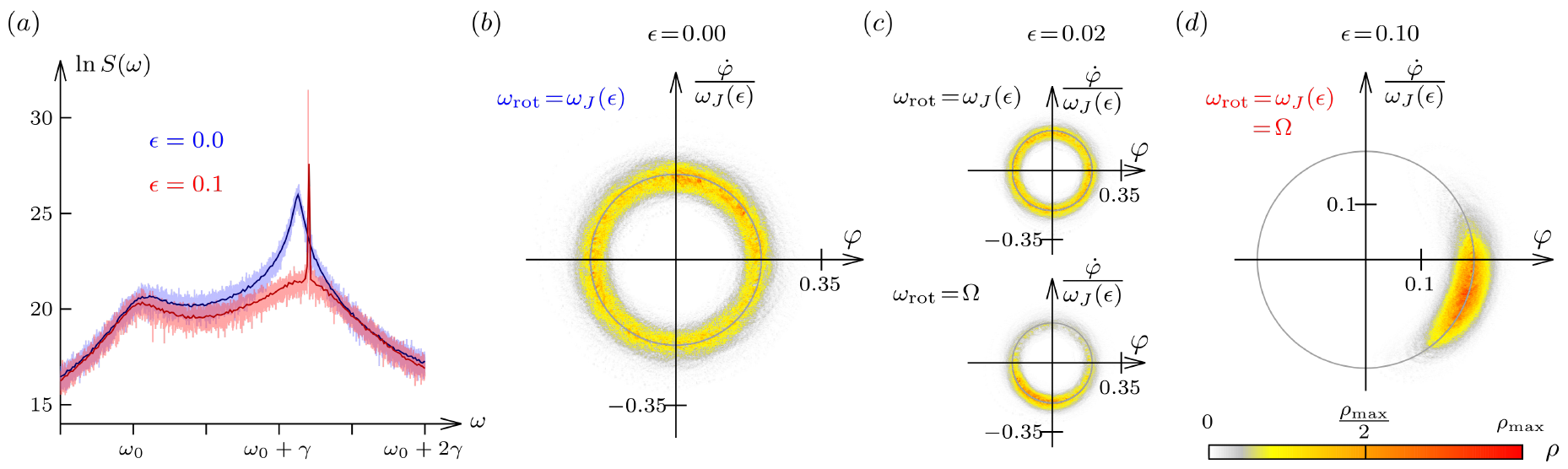}
\caption{(Color online.) Signatures of locking in classical dynamics with noise. (a) The hallmark signature of locking in the emission spectrum is the emergence of an extremely sharp peak (red) at the locking frequency $\Omega$ as compared to the unlocked (thermally) broadened emission (blue) around $\omega_{J}(0)$ [$\Omega= \omega_0 + 1.2 \gamma; \omega_J(0) \approx \omega_0 + 1.13\gamma$, reduced from the dc-bias, $\omega_\text{dc} = \omega_0 + 1.5\gamma$ due to the in-series resistance]. The cavity resonance enhances thermal fluctuations around $\omega_0$. [Light colors show the spectrum averaged over $10$ long runs with $0 \le \omega_0 t \le 10^6$, which is further smoothened  to a resolution $\delta \omega \approx \gamma/80$ 
to give the solid lines. Noise is modeled by an Ornstein-Uhlenbeck process with (zero-frequency) spectral noise density, $S_\xi=\gamma/30$, and correlation time $t_\textrm{corr}=2/\gamma$.] 
Phase space distributions $\rho$ are shown: (b) without a locking signal in a frame rotating with $\omega_J \neq \Omega$; (c) in the unlocked region for frames rotating with $\omega_J$ and with $\Omega$; and (d) for the locked case rotating with $\omega_J = \Omega$. Diffusion around the limit cycle caused by the noise is modulated and eventually confined by increasing the locking signal. In the latter case, long-time phase correlations result in a pronounced reduction of the linewidth shown in (a). [For parameters not otherwise stated, see Fig.~\ref{fig:intro}; (pseudo-) steady-state distributions are gained from a single run with $5 \cdot10^5 \le \omega_0 t \le 10^6$.]
}
\label{fig:lockedspectrum}
\end{center}
\end{figure*}

\section{Effects of noise}\label{sec:noise}

\begin{figure}[t]
\begin{center}
\includegraphics[width=0.99\columnwidth]{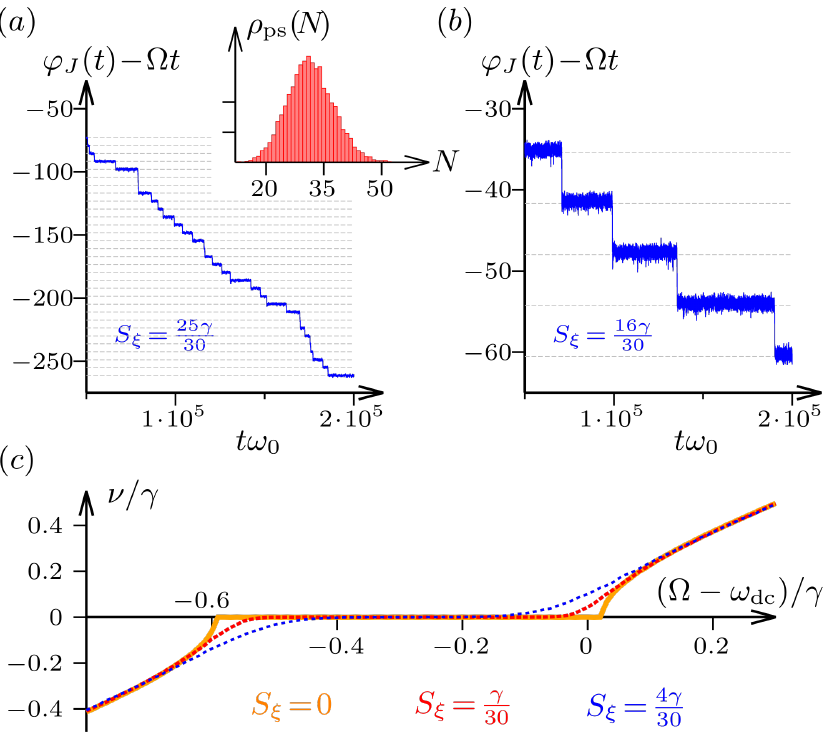}
\caption{(Color online.) Effect of noise on the dynamics of the junction phase.
Noise induced diffusion allows the slow phase variable, $\varphi_J(t)-\Omega t$, to overcome the potential barrier in the tilted washboard potential described by Eq.~\eqref{eq:Adnoise} and slip by $2\pi$ from one potential minimum to the next. An increased noise strength [(a) as compared to (b)] results in a strongly (exponentially) enhanced rate of slips, $\Gamma_\textrm{ps}$, see Eq.~\eqref{eq:slip_rate}. In the simplest regime, slips follow a Poissonian distribution [see histogram of $10^4$ traces in inset of (a)]. The slip rate, $\Gamma_\textrm{ps} := \frac{\langle N \rangle}{\Delta t}= \frac{\langle\!\langle N \rangle\!\rangle}{\Delta t}\approx \frac{30}{1.5\;10^5}\omega_0$ determines the noise induced shift and linewidth of the spectral peak. [Parameters as in Fig.~2(a), $t_\textrm{corr}=2/\gamma$].\\
(c) Bifurcations appearing in locking curves without noise are progressively washed out for increasing noise strength, and even deep in the locked region the average Josephson frequency differs from the locking frequency, $\omega_J(\epsilon)-\Omega \not\equiv 0$, due to (thermal) phase slips. [Parameters cf. Fig.~3(b) top].
}
\label{fig:noise}
\end{center}
\end{figure}

The hallmark of the injection locking phenomenon is that it stabilizes the phase of oscillations, $\psi$, against perturbations due to noise.
To explicitly demonstrate and study this stabilization in numerical simulations stochastic noise forces are added to the equations governing the system's dynamics.

In recent Josephson photonics experiments the dominant source of noise were fluctuations $\Delta V(t)$ of the applied bias voltage which were slow and of classical nature, and ascribed to thermal or external noises.
Another unavoidable source of fluctuations is voltage noise created at the in-series resistance $R_0$ by the shot-noise of the Josephson current across the tunnel junction. This quantum noise, stemming from the granular nature of charges, can only imperfectly be captured by a stochastic force term in a classical equation-of-motion, as we will briefly discuss  further below.

Following the typical experimental situation, we include Gaussian colored noise as an addition to the voltage term in Eqs.~\eqref{eq:eom}.
Before turning again to the effective Adler equation for analytical considerations and estimates, let us discuss the effects of adding noise to Eqs.~\eqref{eq:eom}, as observed in the full numerical solutions.
Fig.~\ref{fig:lockedspectrum}(a) reproduces the hallmark experimental signature of locking from such solutions: the emergence of an extremely sharp peak arising from the (thermally) broadened spectrum of the cavity emission as the system enters the locking region.
The same physics is described by the phase space distributions displayed in Fig.~\ref{fig:lockedspectrum}(b-d). Without a locking signal,  weak and slow fluctuations allow the quadratures of the cavity oscillation to diffuse around the complete limit cycle (with slight excursions in radial direction). A locking signal strongly restrains this diffusion and in the rotating-frame phase space both angle and radius are stabilized in the locked regime, see Fig.~\ref{fig:lockedspectrum}(d).

To understand and analyze the effects of noise observed in the results of the full numerical solutions of Eqs.~\eqref{eq:eom} shown in Fig.~\ref{fig:lockedspectrum},
 we can reconsider the derivation of the effective Adler equation, now in the presence of a fluctuating noise term. Reasoning that the high frequency components of noise fluctuations are efficiently filtered and do not reach the device, as is typically the case in Josephson photonics
experiments, which corresponds to assuming a correlation time of the colored noise larger than typical time-scales of the dynamics, the time-scale separation used in the previous section will remain valid in the presence of noise. It is then easy to show that noise introduces in 
the circuit Adler equation a stochastic frequency detuning due to the variation of the dc bias, 
$\xi(t) = (2e/\hbar)\Delta V(t)$,
\begin{align}
\dot\psi =&\ \nu(\epsilon)  - v_{R_0}\; \frac{\epsilon}{2}\sin(\psi) + \xi(t) = \frac{\partial U(\psi)}{\partial\psi} +\xi(t).\label{eq:Adnoise}
\end{align}

To the overdamped dynamics of the phase particle in the washboard potential $U(\psi) = \nu\psi + \nu_c\cos(\psi)$ tilted by the detuning $\nu$ and with a modulation amplitude increasing with the locking strength, there now contributes a fluctuating noise force, which can alternatively be seen as a fluctuation of the tilt.

In the absence of an injection signal, the potential is a flat tilted landscape, $U(\psi,\epsilon=0) = \nu_0 \psi$; the resultant uniform motion for $\Psi$ with velocity $\nu_0$ corresponds to cavity emission at $\omega_J$. In a phase space rotating with the same frequency $\omega_J$ this motion is represented by a fixed point. Noise adds (free) diffusion to the motion of the phase particle, which yields the wide spectrum for the cavity emission shown by the  blue line of Fig.~\ref{fig:lockedspectrum}(a) and the typical "donut"-shape in the rotating phase-space, Fig.~\ref{fig:lockedspectrum}(b). 
Adding the injection signal provides a series of minima to the potential, separated by $2\pi$, that can 
reduce the effects of noise, restrict the diffusion, and decrease the average velocity.
In a frame rotating with this average velocity [upper panel of Fig.~\ref{fig:lockedspectrum}(c)] one nonetheless obtains diffusion exploring the whole limit cycle and a donut-shaped phase-space distribution. The modulation of this diffusion by the injection signal only becomes apparent in a frame rotating with $\Omega$ [lower panel of Fig.~\ref{fig:lockedspectrum}(c)], where a modulated donut with more or less weight indicating potential minimum and maximum is found. In the locked region, the distribution finally becomes strongly confined, see Fig.~\ref{fig:lockedspectrum}(d). 

Starting from the known tilted washboard dynamics in the absence of noise, we see that adding even weak noise may have strong effects close to the onset of locking around the critical tilt $\nu_c$.  Just above the onset the average velocity of the phase particle $\omega_J-\Omega$ is approaching zero due to the slow creeping motion along the nearly flat part of the washboard. Clearly fluctuations of the tilt will cut this creeping motion short and strongly increase the average velocity, as is indeed seen in Fig.~\ref{fig:noise}(c).
The situation is very different deep inside the locking region, $|\nu/\nu_c|\ll 1$, where, if the spectral density of noise is smaller 
than the barrier height between consecutive minima, the dynamics
of the locked phase particle can be qualitatively described as localized explorations of the local 
minimum (intra-well dynamics) combined with occasional escape events (inter-well dynamics) where the phase
crosses the barrier and is subsequently re-trapped at the next potential minimum. These escape events are noise-assisted phase slips 
where the phase rapidly changes by $2\pi$. Fig.~\ref{fig:noise}(a),(b) depicts the typical time-dependence of $\psi(t)$
in the locked region and under the influence of colored noise generated by an Ornstein-Uhlenbeck process with a correlation time chosen as $2/\gamma$. 
The average rate of phase slips sets 
the average velocity $\omega_J-\Omega$, leading to deviations from the constant part of the locking curve in Fig.~\ref{fig:noise}(c).
The noisy dynamics of the Josephson phase particle $\psi$, described here on the basis of the Adler-like effective equation, is similar to known Shapiro-steps physics, where however, there are no direct equivalents to the signatures in the cavity spectrum and the phase space features. 

In the emerging picture of locking in the presence of noise, the concept of injection locking can no longer be equivalent to $\dot\psi=0$ and must be amended to account for the fact that the relative phase $\psi$ between the injection signal and the oscillations in the Josephson device is no longer time-independent. 

Instead, the stabilization of the phase will be quantified by the rate of phase slips.
For a simple quantitative estimate we turn to the Kramers regime of diffusion over a barrier for overdamped dynamics \cite{Kramers1940}.  Neglecting correlations between consecutive slips and for a noise spectral density $S_\xi\equiv \int d\tau\ \xi(\tau)\xi(0)$  small compared to the barrier height, $U_B$, 
the rate of phase slips $\Gamma_\text{ps}$ is exponentially suppressed, 
\begin{align}\label{eq:slip_rate}
\Gamma_\text{ps}(U_B,S_\xi) \sim 
\exp\left(-\frac{U_B}{S_\xi}\right),\quad \text{for}\  U_B \gg S_\xi.
\end{align}
The barrier height has a maximum, when the detuning in the Adler equation vanishes $\nu(\epsilon)=0$, and where locking is most stable against noise. Its maximal height can be estimated as $U_{B,\text{max}}=\nu_c\simeq eI_cR_0\epsilon/\hbar$. 
This sets an upper limit for the noise intensity that can be overcome
and stabilized by an injection signal with fixed amplitude $\epsilon$,
\begin{align}
S_\xi < (\epsilon/2)v_{R_0}.
\end{align}

The phase stabilization has important consequences for the spectral width of the radiation emitted by the Josephson photonics device.
In absence of an injection signal, the spectral width $\delta\omega$ in the simplest linear scenario is directly set by the noise spectral density, 
$\delta\omega\simeq S_\xi$. By injection locking the device and tuning the injection frequency to the condition of maximum barrier height,
$\nu(\epsilon)=0$, the new spectral width will be given by the rate of
phase slips, $\delta\omega\simeq\Gamma_\text{ps}$. This will amount to an \emph{exponential reduction of the spectral width}, yielding an extremely sharp spectral feature when the noise intensity falls well below the above threshold, as in Fig.~\ref{fig:lockedspectrum}(a). 

The requirements to observe the exponential reduction of the spectral width can be estimated quantitatively based on typical realizations of
the Josephson photonics device in experiments \cite{Hofheinz2011,Rolland2019,Westig2017,Peugeot2020}. For a device with small Josephson coupling $\tilde{I}_c = 2eI_cR/\hbar\omega_0=0.5$,
as can be routinely realized, and a resonance quality factor $Z(\omega=\omega_0)/Z(\omega=0)\approx R/R_0 \simeq 30$ (a low estimate), the barrier height
that can be created with an injection signal of amplitude $\epsilon$ can be estimated to 
$U_{B,\text{max}}\simeq (\epsilon/2)v_{R_0} = eI_cR_0\epsilon/\hbar \simeq (\epsilon/120)\omega_0$.
Furthermore, the spectral width for such a device in absence of an injection signal is typically of the order $\delta\omega\simeq 10^{-3}\omega_0$,
resulting in the same estimation for the noise spectral density $S_\xi\simeq 10^{-3}\omega_0$. Therefore, the requirement to observe 
the exponential suppression of noise $U_{B,\text{max}} > S_\xi$ amounts to $\epsilon > 0.12$. This places the amplitude of the injection signal reasonably well in the linear regime,
 see Fig.~\ref{fig:tongues}.

In the opposite limit, when the injection amplitude is weak, such that the noise
satisfies $S_\xi > (\epsilon/2)v_{R_0}$, the stabilization is negligible [the evolution of the phase
$\psi(t)$ in this case is shown in Fig.~\ref{fig:noise}(a)]. While in this regime the phase is not stabilized and the spectrum remains broad, injection locking may have measurable consequences in other observables, such as in the statistics of the radiation emitted by the device. Interesting questions, such as the relation between 
the statistics of phase slips and the statistics of the emitted radiation, will be addressed elsewhere.

While the estimations above suggest that it may be optimal for locking to increase the resistor $R_0$, this step may become counter-productive if current fluctuations at $R_0$ become the dominant source of voltage noise. 
In that case, thermal noise can become negligible compared to the shot noise, as the Josephson photonics device is operated at low temperature. Therefore, quantum fluctuations
due to Cooper pair tunneling become dominant. 

Conjecturing that one may still arrive at an effective Adler-like dynamical equation, there are obvious ways in which quantum effects will modify the picture of the phase particle in the washboard potential $U(\psi)$. 
The point-like phase particle will be replaced by a wave function, where the scale of (zero-point) quantum fluctuations is associated with a mass assigned to the phase particle, which emerges from the scale of (single Cooper pair) charging effects. Classical thermal diffusion over the barrier is superseded by tunneling across the barrier between consecutive minima of $U(\psi)$ under the influence of dissipation. One may anticipate a particularly interesting regime in this competition between dissipation and tunneling, where the quantum dynamics of the phase particle is described by a Bloch-type wavefunction in the time-crystal defined by potential $U(\psi)$, in spite of dissipative effects. 

Setting up a full model of quantum dynamics of locking in the shot-noise dominated regime to study the quantum statistical properties of the microwave radiation emitted from the Josephson photonics, as well as deriving and studying resulting effective Adler-like equations are intriguing avenues for further research. 
It will link our investigations to interesting recent studies of locking for various systems, where the quantum character of the oscillator becomes crucial \cite{Walter2014,Loerch2016,Loerch2017,Roulet2018,Amitai2018,Koppenhoefer2019,Jessop2020,Lifshitz2021}. 
A further direction is uncovering parallels and distinctions between two different types of (quantum) phase slips; the tunneling of the phase of a dynamical solution, such as the Josephson phase in the injection locked devices studied here, and flux tunneling in thin superconducting wires \cite{Astafiev2012}. 


\section{Higher-order resonances}\label{sec:higher_order}


Part of the attraction of Josephson photonics devices is the variety of resonances associated with multi-photon creation, which are easily addressable in such devices by a simple change of the dc-voltage bias. In addition to the fundamental resonance at $2 e V_\textrm{dc} \approx \hbar \omega_0$, there appear resonances at $2 e V_\textrm{dc} \approx p \hbar \omega_0 \; (p \in \mathbb{N})$, whenever the bias provides each Cooper pair crossing the junction with the energy to excite $p$ photons in the mode $\omega_0$. Such processes were observed already in the very first experiment \cite{Hofheinz2011} and also play a pivotal role for the Josephson laser \cite{Cassidy2017} where, however, they mix with fundamental ($p=1$) resonances for higher modes $\omega_p = p \omega_0$ of the cavity. Equivalent processes in ac-driven Josephson devices \cite{Wustmann2019} have very recently been observed for $p=3$ downconversion \cite{Svenson2017,Svenson2018,Chang2020}.

\begin{figure*}
\begin{center}
\includegraphics[clip,width=0.99\textwidth]{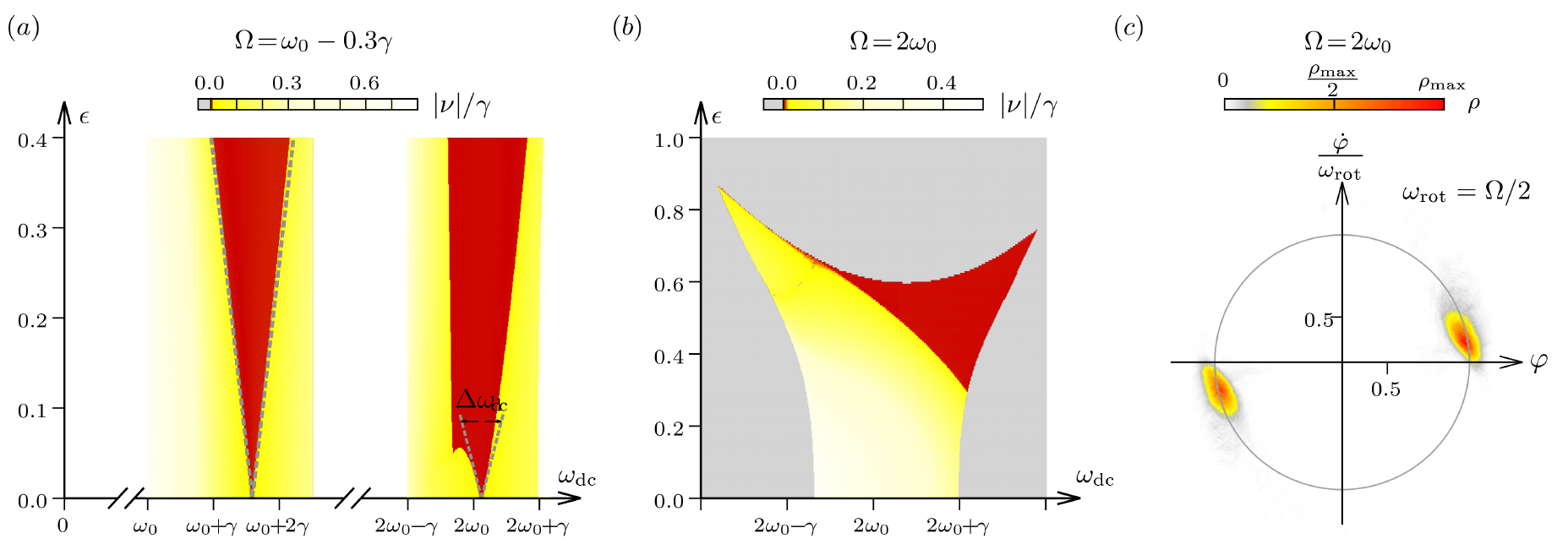}
\caption{(Color online.)  
(a) A locking signal with frequency $\Omega\approx \omega_0$ can lock a Josephson photonics system driven at different dc-biases, $\omega_\textrm{dc}\approx p \Omega$, where each tunneling Cooper pair creates $p$ photons with frequency $\omega_\textrm{dc}/p$. 
The strongly nonlinear driving regime above the threshold of the ($p=2$) parametric resonance, $\tilde{I}_c =2.5 > 2 =: \tilde{I}^\textrm{thr}_c$, modifies the shape of the Arnold tongues. [Dashed lines show analytical results based on an extended theory (non-perturbative in $\tilde{I}_c$, see main text) for both $p=1$ and $p=2$. Parameters are $v_{R_0} = \omega_0/6$ and other parameters as in Fig.~\ref{fig:intro}].
(b) The downconverted oscillations with frequency $\sim \omega_0$ at the ($p=2$) parametric resonance can alternatively be locked by a signal oscillating at $\Omega \approx 2 \omega_0$. The Arnold tongue  [red, defined by $\nu(\epsilon)=[\omega_J(\epsilon) - \Omega]/2 \equiv0$] appears at finite locking amplitude $\epsilon$ and crosses over into regions (grey), where  directly driven oscillations at $\sim 2 \omega_0$ prevail. 
(c) While a locking signal at $\Omega \approx \omega_0$ breaks the degeneracy of parametric oscillations, the degeneracy remains preserved for  $\Omega = 2 \omega_0$ as demonstrated by the phase space distribution  (in a frame rotating with $\Omega/2$)  in the locked region ($\epsilon=0.5$, $\omega_\textrm{dc}=2\omega_0 + \gamma$). Noise ($S_\xi=16\gamma/30$, $t_\textrm{corr}=2/\gamma$) allows the system to slip between the two degenerate solutions, whose symmetry is not broken by the  $\Omega = 2 \omega_0$ locking signal.
}
\label{fig:parametric_fast_lock_sig}
\end{center}
\end{figure*}

In the classical description these processes, which based on a quantum picture may be called p-photon creation processes, materialize as (higher-order) parametric resonances in the equations-of-motion.  
Setting $R_0 \equiv 0$ and $\epsilon=0$ for the moment, Eq.~\eqref{eq:eom2}, reduces to
\begin{align}
\ddot\varphi+\gamma\dot\varphi+\omega_0^2\varphi =& \tilde{I}_c \gamma\omega_0 \sin {(\omega_\textrm{dc} t -\varphi)}\nonumber\\
 \propto& \frac{\varphi^{(p-1)}}{(p-1)!} \sin{ (\omega_\textrm{dc} t + \textrm{const.})  },\label{eq:downconv}
\end{align}
The conventional parametric resonance occurs at $\omega_\textrm{dc} \approx 2 \omega_0$, see 
Refs.~[\onlinecite{Gramich2013,Armour2013,Juha2013,Padurariu2012,Kubala2015,Meister2015,Kubala2020}] for details.

Important for our discussion below are two traits which the conventional parametric ($p=2$) and all higher-order ($p>2$) resonances have in common: (i) Classically, there is a driving threshold above which a solution $\tilde{\varphi}(\omega = \omega_\textrm{dc}/p) = 0$ becomes unstable and a parametric oscillation emerges. 
(ii) The parametric solution with period $T^{(p)} = \frac{2 \pi}{\omega_\textrm{dc} /p} = p T^\textrm{drive}$  is $p$-fold degenerate, with the $p$ solutions connected through time-translation by an integer multiple of the period of the (effective) parametric drive $ T^\textrm{drive} :=  2 \pi/\omega_\textrm{dc}$. 
This corresponds to a $p$-fold symmetry in phase-space [with solutions $\tilde{\varphi}_k(\omega_\textrm{dc}/p) = e^{i 2\pi k/p} \; \tilde{\varphi}_{k=0}(\omega_\textrm{dc}/p)$ with $k=0 \hdots p-1$]. The $p$-fold symmetry has been described in the language of a spontaneous breaking of a discrete symmetry and termed phase-space time-crystal \cite{Guo2013,Zhang2017,Gosner2020,Liang_2018,Nathan2020}.

The physics of phase locking at higher-order resonances is very rich and considerably more complex compared to the fundamental resonance; for one, due to the fact, that the unlocked (classical) solution only exists in a nonlinear regime. Here, we do not want to discuss locking at higher-order resonances at the same level of detail as done for the fundamental resonance above. Instead, we only briefly present first results for $p=2$. These are chosen to highlight a specific locking feature, newly arising for the $p>1$ case, which is of both fundamental interest and practical relevance for possible applications (e.g., for stabilizing a squeezing axis). The locking signal aimed at stabilizing the phase and frequency of the oscillations at the (downconverted) `slow' frequency, $ \omega_\text{dc}/p$, can either \emph{break, or preserve} the discrete $p$-fold time-translation symmetry. The symmetry will be broken by providing a signal at the slow frequency, $\Omega \sim \omega_\textrm{dc}/p \simeq \omega_0$, but preserved by providing a signal at the fast frequency, $\Omega \simeq \omega_\text{dc} \simeq p \omega_0$.

In the first case, the very same locking signal at $\Omega \simeq \omega_0$ can actually lock a Josephson photonics system driven at (two or even multiple) different dc-biases; namely around the fundamental resonance,  $\omega_\text{dc} \simeq \omega_0$, but also around the higher-order resonances, $\omega_\textrm{dc} \simeq p \omega_0$. 
This is demonstrated in Fig.~\ref{fig:parametric_fast_lock_sig}(a), which shows the Arnold tongues of locked cavity oscillations with frequency $\Omega$ for fixed locking signal frequency $\Omega \simeq \omega_0$ and varying dc-bias and locking amplitude. 
Shown are results from the numerical solutions of the full equations-of-motions, Eqs.~\eqref{eq:eom}, for a driving strength above the threshold of ($p=2$) parametric oscillations, compared to analytical results for the locking region's boundaries, based on Adler-like equations for slow variables. For the parametric case, we can make an ansatz analogous to Eqs.~\eqref{eq:phiJs} and \eqref{eq:phis}, and expand around the $\epsilon=0$ solutions determined by nonlinear equations involving Bessel functions \cite{Gramich2013,Armour2013,Meister2015}. 
We find a locking region with width $\Delta \omega_\textrm{dc}$ (as in Fig.~\ref{fig:parametric_fast_lock_sig}(a)) that scales similarly to the fundamental resonance case as $\Delta \omega_\textrm{dc} = A \epsilon v_{R_0}$ with a numerical prefactor $A\approx 1.41$ for the parameters of Fig.~\ref{fig:parametric_fast_lock_sig}(a). [The numerical value of $A$ depends on the nonlinear solutions of equations containing various Bessel-functions in an involved manner, but we find $A\sim {\cal{O}}(1)$ away from bifurcations of the equations.] Dashed lines in Fig.~\ref{fig:parametric_fast_lock_sig}(a) also shows analytical results for the width of the Arnold tongue for $p=1$ obtained by extending Eqs.~\eqref{eq:slow1}-\eqref{eq:slow3} for larger driving $\tilde{I}_c$ and expanding in $\epsilon$ the resulting Bessel function expressions. For this case $p=1$, the width of the Arnold tongue is given by $ \Delta \omega_\textrm{dc} = B \epsilon v_{R_0}$ where $B\approx 0.626$ for the parameters of Fig.~\ref{fig:parametric_fast_lock_sig}(a). \\
Clearly, the mixing of downconverted and direct drive, and the nonlinearity of the $\epsilon=0$ solution limits the validity of this ansatz more severely for $p \ge 2$ than for the $p=1$ case, and capturing all features of Fig.~\ref{fig:parametric_fast_lock_sig}(a) is beyond its scope.

For the second case, where the locking signal is provided at the fast frequency, $\Omega \approx \omega_\textrm{dc} \approx 2 \omega_0$, Fig.~\ref{fig:parametric_fast_lock_sig}(b) reveals that a large locking amplitude is required even for optimal detuning, i.e., the Arnold tongue does not touch the $\epsilon =0$ axis.
Intuitively one may argue, that to lock oscillations at the slow frequency, $\Omega/p \sim \omega_0$, a locking signal provided at the fast frequency, $\Omega \sim \omega_\textrm{dc} \sim p \omega_0$, has to be downconverted, which only becomes effective above a certain threshold of signal strength $\epsilon$. Moreover, however, there will be competition between direct locking of the cavity oscillations to a frequency $\Omega$ and the desired parametric locking to $\Omega/p$. The extended Adler-like effective description of parametric locking that could explain the details of Fig.~\ref{fig:parametric_fast_lock_sig}(b) remains a subject for further studies. What is known and strikingly demonstrated by Fig.~\ref{fig:parametric_fast_lock_sig}(c) is the crucial coveted feature of parametric locking: the preservation of the $p$-fold phase-space symmetry. As can be seen directly from the equations-of-motions, if a solution, where the cavity is locked at the slow frequency, $\Omega/p$, exists, it is degenerate and there exist $p$-equivalent solutions for the cavity oscillations shifted by an integer multiple of $2 \pi/\Omega$ in time, or rotated by an integer multiple of $2 \pi/p$ in phase space. In the phase-space distribution of  Fig.~\ref{fig:parametric_fast_lock_sig}(c)  the $2$-fold symmetry is clearly seen. In presence of noise as discussed in Sec.~\ref{sec:noise}, the cavity oscillations will explore both possible degenerate solutions.

We expect that the scenario of the parametric locking that preserves symmetry will also be reproduced in a quantum description. This mechanism may be used, for instance, to reduce the diffusion of the orientation of the squeezing axis for degenerate or non-degenerate emission of quantum microwave radiation from dc-biased Josephson photonics devices without modifying other desired quantum emission properties. It may also considerably simplify experiments such as the recent confirmation of entanglement in Ref.~[\onlinecite{Peugeot2020}] and enable new applications. Note, that a recent experiment on non-degenerate parametric oscillations \cite{Svenson2018} follows the diametrically opposite approach of applying a weak on-resonant tone as described above and consequently breaks the phase-space symmetry.


\section{Synchronization}\label{sec:synch}


\begin{figure}[t]
\begin{center}
\includegraphics[width=0.99\columnwidth]{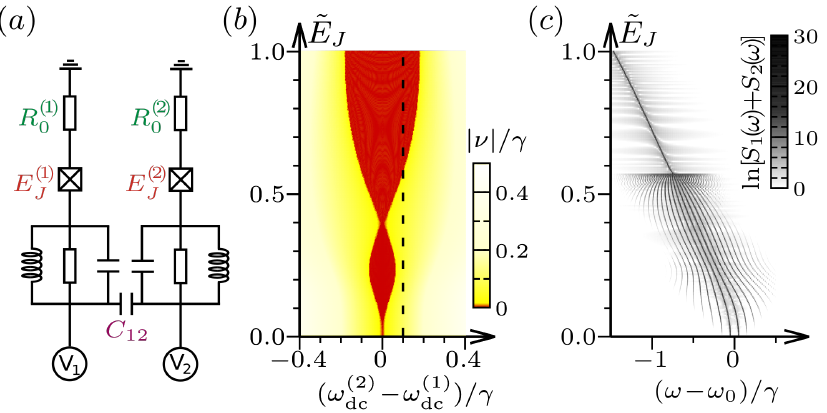}
\caption{(Color online.) Synchronization of Josephson oscillators. 
(a) Sketch of two lockable Josephson photonics circuits, which can synchronize if weakly coupled (e.g., by a capacitance). 
(b) Arnold tongue indicating the voltage detuning between the two circuits, where oscillations synchronize [red, $\nu(\epsilon) =  (\omega_J^{(2)} - \omega_J^{(1)})/\gamma$], as the effective coupling increases with $\tilde{E}_J$. The coupling terms in the Adler-Kuramoto equations \eqref{eqMT:lin} scale with the critical detuning parameters \eqref{eq:crit_detun}, $\nu^{(\sigma)}_c \propto  (\tilde{E}_J)^2$, explaining the shape of the tongue for weak coupling, while the change of Josephson frequencies with $\tilde{E}_J$ yields more complex behavior (such as multiple (de)synchronization transitions) at larger $\tilde{E}_J$. [results for identical cavities ($\tilde{E}_J^{(1)}= \tilde{E}_J^{(2)}=\tilde{E}_J$, etc.) 
with anti-symmetric bias, $\omega_\textrm{dc}^{(2/1)}= \omega_0 \pm (\omega_\textrm{dc}^{(2)} -  \omega_\textrm{dc}^{(1)} )$ and  parameters as in Fig.~\ref{fig:intro}; $\epsilon^{(\sigma)}=C_{12}/(C^{(\sigma)} + C_{12})=0.01$.]
(c) (Sum) spectrum at bias $\omega_{dc}^{(2/1)} = \omega_0 \pm  \gamma/20$ [dashed in (b)] showing frequency pulling and a fan of frequencies before both cavities synchronize to emit with a  common single frequency for large $\tilde{E}_J$.
}
\label{fig:synch}
\end{center}
\end{figure}

The phenomenon of synchronization describes the mutual phase locking of weakly coupled self-sustained oscillators \cite{Kuramoto1975,Pikovsky2001}.
We envision a situation where two Josephson devices, each modeled by a circuit such as studied in the previous sections, 
are weakly coupled. 
Experimentally a capacitive, inductive or even resistive coupling can be easily realized and potentially tunable (or nonlinear) coupling scenarios could also be engineered by linking the two devices by a Josephson junction or more complex circuit.

The universal nature of synchronization implies that the resulting mutually locked dynamics are qualitatively similar for any of these type of couplings between the devices (even though variations may be envisioned for time-delayed or strongly nonlinear couplings).
To provide an example of the dynamics, we have chosen to describe
two Josephson circuits coupled by a mutual capacitance $C_{12}$, as depicted in Fig.~\ref{fig:synch}, assumed small compared to the capacitance of the circuits
$C_{12}\ll C^{(1)},C^{(2)}$. We will demonstrate that the capacitive
coupling leads to synchronization of the Josephson photonics devices. However, the formalism presented, as well as the qualitative aspects of our
discussion, apply to any coupling mechanism.

\begin{table}[t]
\centering
\begin{tabular}{l @{\hskip 0.3cm} c @{\hskip 0.3cm} l} 
 \hline
 Physical Quantity & Parametrization & Unit \\ [0.5ex] 
 \hline
 \vspace*{0.1cm}
 resonance frequencies & $\omega_0^{(\sigma)}$ & s$^{-1}$ \\
 \vspace*{0.1cm}
 resonance widths    & $\gamma^{(\sigma)}=\left(R^{(\sigma)}C^{(\sigma)}\right)^{-1}$ & s$^{-1}$\\
 \vspace*{0.1cm}
 d.c. bias voltages & $\omega^{(\sigma)}_\text{dc}=(2e/\hbar) V^{(\sigma)}$ & s$^{-1}$ \\
 \vspace*{0.1cm}
 Josephson frequencies & $\omega_J^{(\sigma)}\neq \omega^{(\sigma)}_\text{dc}$ & s$^{-1}$ \\
 \vspace*{0.1cm}
 scale of residual voltages & $v^{(\sigma)}_{R_0}=(2e/\hbar) I^{(\sigma)}_c R^{(\sigma)}_0$ & s$^{-1}$ \\ [1ex]
 \vspace*{0.1cm}
 Josephson driving strengths  & $\tilde{I}_c^{(\sigma)} = (2e/\hbar)I^{(\sigma)}_cR^{(\sigma)}/\omega^{(\sigma)}_0$ & $1$ \\
 coupling strenghts & $\epsilon^{(\sigma)}=C_{12}/\left(C^{(\sigma)}+C_{12}\right)$ & 1 \\
 \hline
\end{tabular}
\caption{Physical quantities that characterize the two devices, $\sigma=1,2$.}
\label{table:2}
\end{table}

The system in Fig.~\ref{fig:synch} is governed by the following (Kirchhoff-)equations of motion
\begin{subequations}\label{eq:EOM-synch}
\begin{align}
\dot\varphi_{J}^{(\sigma)} &= \omega_\text{dc}^{(\sigma)} - v_{R_0}^{(\sigma)}\sin\left(\varphi_{J}^{(\sigma)}\right) - \dot\varphi^{(\sigma)} \label{eq:J}\\
Z^{(\sigma)}\left[\varphi^{(\sigma)}\right] &= \tilde I_{c}^{(\sigma)}\sin\left(\varphi_{J}^{(\sigma)}\right)+\frac{\epsilon^{(\sigma)}}{\omega_{0}^{(\sigma)}\gamma^{(\sigma)}}\ddot\varphi^{(\bar{\sigma})}\label{eq:cav}
\end{align}
\end{subequations}
where we have introduced superscripts $\sigma=1,2$ to label the two devices and $\bar{\sigma}$, defined as $\bar{\sigma}=2$ if $\sigma=1$, $\bar{\sigma}=1$ if $\sigma=2$.

The dimensionless response $Z^{(\sigma)}$ that describes the two cavities is defined by 
\begin{align}
Z^{(\sigma)}\left[\varphi^{(\sigma)}\right] = \frac{1}{\omega^{(\sigma)}_0\gamma^{(\sigma)}} \left[\ddot \varphi^{(\sigma)}+\gamma^{(\sigma)}\dot \varphi^{(\sigma)}+\left(\omega_{0}^{(\sigma)}\right)^2\varphi^{(\sigma)}\right],
\end{align}
and further notation is summarized in Table~\ref{table:2}.

In close analogy to our calculations for locking in Sec. \ref{sec:injection} we can now derive an effective equation for the slow dynamics describing how and in which parameter region the two systems synchronize. The synchronization region corresponds to a range of detuning values between the two dc voltages, 
$\omega^{(2)}_\text{dc}-\omega^{(1)}_\text{dc}$, 
where due to the coupling $\epsilon$ the frequency of Josephson oscillations will be identical in both circuits.  To derive the Kuramoto-Adler equations, from which the boundaries of the synchronization region follow, we introduce slowly varying functions, as in Sec. \ref{sec:injection}, for the Josephson and cavity phases, 
\begin{align}
\varphi_{J}^{(\sigma)} =&\ \omega_{J}^{(\sigma)} + \theta_J^{(\sigma)}(t)+a^{(\sigma)} \sin\left[\omega_{J}^{(\sigma)}t+\phi_{J}^{(\sigma)}(t)\right],\label{eq:ansatz_sync_JJ}\\
\varphi^{(\sigma)} =&\ b^{(\sigma)}\sin\left[\omega_{J}^{(\sigma)}t+\phi^{(\sigma)}(t)\right].\label{eq:ansatz_sync_cav}
\end{align}
The ansatz above is consistent with the limit of weak driving $\tilde I_{c}^{(\sigma)}\ll 1$ and small low-frequency impedance $R_0^{(\sigma)}\ll R^{(\sigma)}$. Assuming time scale separation and further approximations (see S.M. for details) a  system of coupled equations for the slowly-varying functions can be derived which finally reduces for weak capacitative coupling  to the Kuramoto model
\begin{subequations}\label{eqMT:lin}
\begin{align}
\!\!\!\! \dot\theta_J^{(1)} \!=\!&\  \tilde\nu^{(1)}+\epsilon_1 \nu_c^{(1)}  \sin [ \nu_Jt+\theta_J^{(2)}-\theta_J^{(1)}-\chi^{(1)}-\chi^{(2)}]  \label{eqMT:lin1}\\
\!\!\!\!  \dot\theta_J^{(2)} \!=\!&\ \tilde\nu^{(2)}-\epsilon_2 \nu_c^{(2)}  \sin[\nu_Jt+\theta_J^{(2)}-\theta_J^{(1)}+\chi^{(1)}+\chi^{(2)}]. \label{eqMT:lin2}
\end{align}
\end{subequations}
Here, $\tilde\nu^{(\sigma)}\equiv \omega_J^{(\sigma)}(\epsilon=0)-\omega_J^{(\sigma)}$ denotes the detuning between 
 the Josephson frequency in absence of coupling, $\omega_J^{(\sigma)}(\epsilon=0)$, i.e. at $C_{12}=0$, and the Josephson frequency $\omega_J$ affected by mutual frequency pulling. 
 The two critical detuning parameters $\nu_c^{i}$ are found to be
\begin{subequations}\label{eq:crit_detun} 
\begin{align}
\nu_c^{(1)} =&\  \frac{1}{2}v_{R_0}^{(1)}\frac{\omega_J^{(1)}}{\gamma^{(2)}}\frac{\tilde I_{c}^{(2)}}{|z^{(1)}||z^{(2)}|},\label{eq:nuc1}\\
\nu_c^{(2)} =&\ \frac{1}{2}v_{R_0}^{(2)}\frac{\omega_J^{(2)}}{\gamma^{(1)}}\frac{\tilde I_{c}^{(1)}}{|z^{(1)}||z^{(2)}|}. \label{eq:nuc2}
\end{align}
\end{subequations}
Both set of parameters depend on complex dimensionless impedances 
$z^{(\sigma)}\equiv |z^{(\sigma)}| e^{i\chi^{(\sigma)}}=\left(2\nu^{(\sigma)}+i\gamma^{(\sigma)}\right)/\gamma^{(\sigma)}$, that are obtained from the Fourier transform of the responses $Z^{(\sigma)}$ evaluated at the detuning between the corresponding Josephson frequency and the resonance (see S.M.).
This forms an implicit set of equations together with the definition of the Josephson frequency [based on averaging Eq.~\eqref{eq:J}]
\begin{align}
\omega_J^{(\sigma)} = \omega_\text{dc}^{(\sigma)} - \frac{1}{2}\frac{\left(v_{R_0}^{(\sigma)}\right)^2}{\omega_{0}^{(\sigma)}} 
+ \frac{1}{2}\frac{\tilde I_{c}^{(\sigma)}}{|z^{(\sigma)}|}v_{R_0}^{(\sigma)}\sin\left(\chi^{(\sigma)}\right)\,.
\end{align}
Notably, the Josephson frequency without coupling $\omega^{(\sigma)}_J(\epsilon=0)$ depends only on the parameters of the corresponding device $\sigma=1,2$.

Of particular interest in Eq.~\eqref{eqMT:lin} is the appearance of the phases $\chi^{(\sigma)}$ of the oscillators' impedances. These reflect the indirect nature of the coupling of the synchronized Josephson junction phases. Building on the notion, that synchronization can be understood as one junction locking onto a signal provided by the other junction's oscillation, we can immediately discern from Fig.~\ref{fig:synch}(a) that the locking signal is mediated by the cavities. The signal provided by the first JJ's oscillation driving its cavity thus comes with an extra phase $\chi^{(1)}$ determined by that cavity's response function. That signal, in turn, drives the second circuit via an oscillating current, i.e. in a manner slightly different from the oscillating voltage drive considered in Secs.~\ref{sec:pre} and \ref{sec:injection}. Considering the corresponding changes to the circuit equations \eqref{eq:eom} and the resulting Eqs.\eqref{eq:slow1}-\eqref{eq:slow3}, one notes that the cavity response enters again and brings an extra phase of $\chi^{(2)}$, thus explaining the phase shifts in  Eq.~\eqref{eqMT:lin}. Also interesting to note considering Eq.~\eqref{eqMT:lin} is the fact that the effective coupling can be amplified by a dimensionless factor, $\omega_J^{(\sigma)}/\gamma^{(\bar{\sigma})}$,  similar to the quality factors of the cavities. 
Therefore, a significant coupling can be achieved between 
high quality factor Josephson devices despite a relatively low value of the mutual capacitance.
 
To emphasize the connection to an Adler-type equation, we define the relative phase of the Josephson oscillations
$\psi \equiv \tilde\varphi_{J}^{(2)}-\tilde\varphi_{J}^{(1)} = \nu_J t + \theta_{J}^{(2)}-\theta_{J}^{(1)}$, where we introduced slow Josephson phases $\tilde\varphi_{J}^{(\sigma)} = \omega_{J}^{(\sigma)} t + \theta_{J}^{(\sigma)}$, obtained from the total Josephson phase, Eq.~ \eqref{eq:ansatz_sync_JJ}, by averaging over
timescales of the order of $2\pi/\omega_J^{(\sigma)}$. 
With this definition, the condition of synchronization corresponds to $\dot\psi = 0$. The difference of the two Adler-Kuramoto equations
yields an Adler-type equation for $\psi(t)$,
\begin{align}
\dot\psi = \tilde\nu^{(2)}-\tilde\nu^{(1)}-
\nu_\textrm{eff} \sin\left(\psi+\chi_\textrm{eff}\right),
\label{synchAdler}
\end{align}
with the effective critical detuning $\nu_\textrm{eff}$ and effective phase offset $\chi_\textrm{eff}$ given by
\begin{align}
\nu_\textrm{eff} =& \sqrt{\left(\epsilon^{(1)} \nu_c^{(1)}\right)^2+\left(\epsilon^{(2)} \nu_c^{(2)}\right)^2};\\
\chi_\textrm{eff} =& \arctan\left[\frac{\epsilon^{(2)} \nu_c^{(2)}-\epsilon^{(1)} \nu_c^{(1)}} 
{\epsilon^{(1)} \nu_c^{(1)} +\epsilon^{(2)} \nu_c^{(2)}}
\tan\left(\chi^{(1)}+\chi^{(2)}\right)
\right].\nonumber
\end{align}
Fig.~\ref{fig:synch}(b) illustrates the Arnold tongue described by the effective Adler equation, Eq.~\eqref{synchAdler}, where we ramp up the effective coupling by increasing the Josephson driving strengths, assumed equal for the two devices.The possibility to achieve synchronization by tuning the Josephson driving strength (using SQUIDs) may be experimentally more immediately feasible than direct control of the coupling capacitance.
At low coupling, the synchronization window increases proportional to the square of the Josephson driving strength, $\nu_\textrm{eff}\propto \tilde{I}_c^2$, consistent with the behavior predicted by Eq.~\eqref{eq:crit_detun}.
Remarkably, the synchronization window is not increasing monotonously with the Josephson driving strength, suggesting the experiment may exhibit sweet spots where synchronization is more efficient. 
Examining 
the analytical expressions, the behavior can be attributed to the dependence of the effective coupling on the dimensionless impedances $z_1$ and $z_2$ of the two coupled cavities. Due to frequency pulling, the detunings $\nu^{(\sigma)}$ may pass through their corresponding resonance ($\nu^{(\sigma)}=0$) at different values of the Josephson driving strength, giving rise to a complicated dependence of the effective coupling.

\section{Conclusions}\label{sec:concl}
In this paper, we have studied injection locking and synchronization in Josephson photonics devices in the classical regime. 
We found that a residual resistance in-series to the Josephson junction-cavity circuit is a crucial ingredient. If it is accounted for, a single-mode circuit weakly driven at its fundamental resonance undergoes self-sustained oscillations and therefore constitutes the simplest Josephson photonics device that can be injection locked and synchronized. 

Based on the fundamental Kirchhoff circuit equations we derived an Adler-type equation describing locking to an additional ac-voltage by  a timescale separation ansatz. The predicted scaling of the locking region with the device parameters will allow experimentalists to devise optimal circuit designs. Bounds on electrical noises against which the phase of cavity oscillations can be stabilized were derived and the noise dependence of the (strongly reduced) linewidth of the cavity emission were discussed. 

At a parametric dc-voltage drive $\omega_{\text{dc}} \approx 2\omega_0$, the downconverted oscillations can be locked with an ac signal being either at the parametric driving frequency, $\Omega \approx \omega_{\text{dc}}$, or at the downconverted frequency, $\Omega \approx \omega_{\text{dc}}/2$. These two scenarios correspond to breaking and preserving of the two-fold time-translation symmetry of the cavity oscillations. In the latter case, noise allows for slips between the two degenerate solutions, yielding a phase-space distribution with two meta-stable solutions in the locked steady state. 
The phase stabilization of Josephson photonics devices by locking will boost their potential as sources of entangled and squeezed quantum microwaves.  

Finally, we extended the model to study synchronization between two Josephson photonics devices. An analytical derivation of a Kuramoto-Adler-type equation again allows quantitative statements on the parameter dependence of synchronization regions, identified by characteristic emission spectra and easily mapped out by tuning the Josephson energy of the two devices.

While in this work the dynamics of the Josephson photonics circuits is governed by classical circuit equations and we studied stability against classical noise, the generics features of the locking and synchronization mechanism, in particular, the importance of an in-series resistance, are expected to carry over to a considerable extent to current experimental devices. These can be designed or tuned to operate in regimes where the dynamics are more or less strongly affected by quantum fluctuations, so that Josephson photonics devices may allow for a systematic study of locking and synchronization from the semiclassical to the deep quantum regime. 
A full theoretical quantum mechanical description including the residual resistance will be required to properly describe regimes where shot noise becomes more dominant than thermal voltage noise and to study such fascinating problems as quantum slips of the phase of the emitted light.

\section*{Acknowledgments}
We aknowledge fruitful discussions with Andrew Armour, Benjamin Huard, and Simon Dambach.

The authors acknowledge funding through the DFG grant AN336/13-1, the Carl Zeiss foundation, and the Center for Integrated Quantum Science and Technology ($\text{IQ}^\text{ST}$).
The authors acknowledge support by the state of Baden-Württemberg through bwHPC
and the German Research Foundation (DFG) through grant no INST 40/575-1 FUGG (JUSTUS 2 cluster).

\bibliography{journalabbreviations,references} 

\begin{thebibliography}{56}%
\makeatletter
\providecommand \@ifxundefined [1]{%
 \@ifx{#1\undefined}
}%
\providecommand \@ifnum [1]{%
 \ifnum #1\expandafter \@firstoftwo
 \else \expandafter \@secondoftwo
 \fi
}%
\providecommand \@ifx [1]{%
 \ifx #1\expandafter \@firstoftwo
 \else \expandafter \@secondoftwo
 \fi
}%
\providecommand \natexlab [1]{#1}%
\providecommand \enquote  [1]{``#1''}%
\providecommand \bibnamefont  [1]{#1}%
\providecommand \bibfnamefont [1]{#1}%
\providecommand \citenamefont [1]{#1}%
\providecommand \href@noop [0]{\@secondoftwo}%
\providecommand \href [0]{\begingroup \@sanitize@url \@href}%
\providecommand \@href[1]{\@@startlink{#1}\@@href}%
\providecommand \@@href[1]{\endgroup#1\@@endlink}%
\providecommand \@sanitize@url [0]{\catcode `\\12\catcode `\$12\catcode
  `\&12\catcode `\#12\catcode `\^12\catcode `\_12\catcode `\%12\relax}%
\providecommand \@@startlink[1]{}%
\providecommand \@@endlink[0]{}%
\providecommand \url  [0]{\begingroup\@sanitize@url \@url }%
\providecommand \@url [1]{\endgroup\@href {#1}{\urlprefix }}%
\providecommand \urlprefix  [0]{URL }%
\providecommand \Eprint [0]{\href }%
\providecommand \doibase [0]{http://dx.doi.org/}%
\providecommand \selectlanguage [0]{\@gobble}%
\providecommand \bibinfo  [0]{\@secondoftwo}%
\providecommand \bibfield  [0]{\@secondoftwo}%
\providecommand \translation [1]{[#1]}%
\providecommand \BibitemOpen [0]{}%
\providecommand \bibitemStop [0]{}%
\providecommand \bibitemNoStop [0]{.\EOS\space}%
\providecommand \EOS [0]{\spacefactor3000\relax}%
\providecommand \BibitemShut  [1]{\csname bibitem#1\endcsname}%
\let\auto@bib@innerbib\@empty
\bibitem [{\citenamefont {Liu}\ \emph {et~al.}(2015)\citenamefont {Liu},
  \citenamefont {Stehlik}, \citenamefont {Eichler}, \citenamefont {Gullans},
  \citenamefont {Taylor},\ and\ \citenamefont {Petta}}]{Liu2015}%
  \BibitemOpen
  \bibfield  {author} {\bibinfo {author} {\bibfnamefont {Y.-Y.}\ \bibnamefont
  {Liu}}, \bibinfo {author} {\bibfnamefont {J.}~\bibnamefont {Stehlik}},
  \bibinfo {author} {\bibfnamefont {C.}~\bibnamefont {Eichler}}, \bibinfo
  {author} {\bibfnamefont {M.~J.}\ \bibnamefont {Gullans}}, \bibinfo {author}
  {\bibfnamefont {J.~M.}\ \bibnamefont {Taylor}}, \ and\ \bibinfo {author}
  {\bibfnamefont {J.~R.}\ \bibnamefont {Petta}},\ }\href {\doibase
  10.1126/science.aaa2501} {\bibfield  {journal} {\bibinfo  {journal}
  {Science}\ }\textbf {\bibinfo {volume} {347}},\ \bibinfo {pages} {285}
  (\bibinfo {year} {2015})},\ \Eprint
  {http://arxiv.org/abs/https://science.sciencemag.org/content/347/6219/285.full.pdf}
  {https://science.sciencemag.org/content/347/6219/285.full.pdf} \BibitemShut
  {NoStop}%
\bibitem [{\citenamefont {Markovi\ifmmode~\acute{c}\else \'{c}\fi{}}\ \emph
  {et~al.}(2019)\citenamefont {Markovi\ifmmode~\acute{c}\else \'{c}\fi{}},
  \citenamefont {Pillet}, \citenamefont {Flurin}, \citenamefont {Roch},\ and\
  \citenamefont {Huard}}]{Huard2019}%
  \BibitemOpen
  \bibfield  {author} {\bibinfo {author} {\bibfnamefont {D.}~\bibnamefont
  {Markovi\ifmmode~\acute{c}\else \'{c}\fi{}}}, \bibinfo {author}
  {\bibfnamefont {J.}~\bibnamefont {Pillet}}, \bibinfo {author} {\bibfnamefont
  {E.}~\bibnamefont {Flurin}}, \bibinfo {author} {\bibfnamefont
  {N.}~\bibnamefont {Roch}}, \ and\ \bibinfo {author} {\bibfnamefont
  {B.}~\bibnamefont {Huard}},\ }\href {\doibase
  10.1103/PhysRevApplied.12.024034} {\bibfield  {journal} {\bibinfo  {journal}
  {Phys. Rev. Applied}\ }\textbf {\bibinfo {volume} {12}},\ \bibinfo {pages}
  {024034} (\bibinfo {year} {2019})}\BibitemShut {NoStop}%
\bibitem [{\citenamefont {Cassidy}\ \emph {et~al.}(2017)\citenamefont
  {Cassidy}, \citenamefont {Bruno}, \citenamefont {Rubbert}, \citenamefont
  {Irfan}, \citenamefont {Kammhuber}, \citenamefont {Schouten}, \citenamefont
  {Akhmerov},\ and\ \citenamefont {Kouwenhoven}}]{Cassidy2017}%
  \BibitemOpen
  \bibfield  {author} {\bibinfo {author} {\bibfnamefont {M.~C.}\ \bibnamefont
  {Cassidy}}, \bibinfo {author} {\bibfnamefont {A.}~\bibnamefont {Bruno}},
  \bibinfo {author} {\bibfnamefont {S.}~\bibnamefont {Rubbert}}, \bibinfo
  {author} {\bibfnamefont {M.}~\bibnamefont {Irfan}}, \bibinfo {author}
  {\bibfnamefont {J.}~\bibnamefont {Kammhuber}}, \bibinfo {author}
  {\bibfnamefont {R.~N.}\ \bibnamefont {Schouten}}, \bibinfo {author}
  {\bibfnamefont {A.~R.}\ \bibnamefont {Akhmerov}}, \ and\ \bibinfo {author}
  {\bibfnamefont {L.~P.}\ \bibnamefont {Kouwenhoven}},\ }\href {\doibase
  10.1126/science.aah6640} {\bibfield  {journal} {\bibinfo  {journal}
  {Science}\ }\textbf {\bibinfo {volume} {355}},\ \bibinfo {pages} {939}
  (\bibinfo {year} {2017})},\ \Eprint
  {http://arxiv.org/abs/https://science.sciencemag.org/content/355/6328/939.full.pdf}
  {https://science.sciencemag.org/content/355/6328/939.full.pdf} \BibitemShut
  {NoStop}%
\bibitem [{\citenamefont {Chen}\ \emph
  {et~al.}(2014{\natexlab{a}})\citenamefont {Chen}, \citenamefont {Li},
  \citenamefont {Armour}, \citenamefont {Brahimi}, \citenamefont {Stettenheim},
  \citenamefont {Sirois}, \citenamefont {Simmonds}, \citenamefont {Blencowe},\
  and\ \citenamefont {Rimberg}}]{Chen2014}%
  \BibitemOpen
  \bibfield  {author} {\bibinfo {author} {\bibfnamefont {F.}~\bibnamefont
  {Chen}}, \bibinfo {author} {\bibfnamefont {J.}~\bibnamefont {Li}}, \bibinfo
  {author} {\bibfnamefont {A.~D.}\ \bibnamefont {Armour}}, \bibinfo {author}
  {\bibfnamefont {E.}~\bibnamefont {Brahimi}}, \bibinfo {author} {\bibfnamefont
  {J.}~\bibnamefont {Stettenheim}}, \bibinfo {author} {\bibfnamefont
  {A.}~\bibnamefont {Sirois}}, \bibinfo {author} {\bibfnamefont {R.~W.}\
  \bibnamefont {Simmonds}}, \bibinfo {author} {\bibfnamefont {M.~P.}\
  \bibnamefont {Blencowe}}, \ and\ \bibinfo {author} {\bibfnamefont {A.~J.}\
  \bibnamefont {Rimberg}},\ }\href {\doibase
  https://doi.org/10.1103/PhysRevB.90.020506} {\bibfield  {journal} {\bibinfo
  {journal} {Phys. Rev. B}\ }\textbf {\bibinfo {volume} {90}},\ \bibinfo
  {pages} {020506} (\bibinfo {year} {2014}{\natexlab{a}})}\BibitemShut
  {NoStop}%
\bibitem [{\citenamefont {Hofheinz}\ \emph {et~al.}(2011)\citenamefont
  {Hofheinz}, \citenamefont {Portier}, \citenamefont {Baudouin}, \citenamefont
  {Joyez}, \citenamefont {Vion}, \citenamefont {Bertet}, \citenamefont
  {Roche},\ and\ \citenamefont {Est{\`e}ve}}]{Hofheinz2011}%
  \BibitemOpen
  \bibfield  {author} {\bibinfo {author} {\bibfnamefont {M.}~\bibnamefont
  {Hofheinz}}, \bibinfo {author} {\bibfnamefont {F.}~\bibnamefont {Portier}},
  \bibinfo {author} {\bibfnamefont {Q.}~\bibnamefont {Baudouin}}, \bibinfo
  {author} {\bibfnamefont {P.}~\bibnamefont {Joyez}}, \bibinfo {author}
  {\bibfnamefont {D.}~\bibnamefont {Vion}}, \bibinfo {author} {\bibfnamefont
  {P.}~\bibnamefont {Bertet}}, \bibinfo {author} {\bibfnamefont
  {P.}~\bibnamefont {Roche}}, \ and\ \bibinfo {author} {\bibfnamefont
  {D.}~\bibnamefont {Est{\`e}ve}},\ }\href {\doibase
  https://doi.org/10.1103/PhysRevLett.106.217005} {\bibfield  {journal}
  {\bibinfo  {journal} {Phys. Rev. Lett.}\ }\textbf {\bibinfo {volume} {106}},\
  \bibinfo {pages} {217005} (\bibinfo {year} {2011})}\BibitemShut {NoStop}%
\bibitem [{\citenamefont {Grimm}\ \emph {et~al.}(2019)\citenamefont {Grimm},
  \citenamefont {Blanchet}, \citenamefont {Albert}, \citenamefont
  {Lepp\"akangas}, \citenamefont {Jebari}, \citenamefont {Hazra}, \citenamefont
  {Gustavo}, \citenamefont {Thomassin}, \citenamefont {Dupont-Ferrier},
  \citenamefont {Portier},\ and\ \citenamefont {Hofheinz}}]{Grimm2019}%
  \BibitemOpen
  \bibfield  {author} {\bibinfo {author} {\bibfnamefont {A.}~\bibnamefont
  {Grimm}}, \bibinfo {author} {\bibfnamefont {F.}~\bibnamefont {Blanchet}},
  \bibinfo {author} {\bibfnamefont {R.}~\bibnamefont {Albert}}, \bibinfo
  {author} {\bibfnamefont {J.}~\bibnamefont {Lepp\"akangas}}, \bibinfo {author}
  {\bibfnamefont {S.}~\bibnamefont {Jebari}}, \bibinfo {author} {\bibfnamefont
  {D.}~\bibnamefont {Hazra}}, \bibinfo {author} {\bibfnamefont
  {F.}~\bibnamefont {Gustavo}}, \bibinfo {author} {\bibfnamefont {J.-L.}\
  \bibnamefont {Thomassin}}, \bibinfo {author} {\bibfnamefont {E.}~\bibnamefont
  {Dupont-Ferrier}}, \bibinfo {author} {\bibfnamefont {F.}~\bibnamefont
  {Portier}}, \ and\ \bibinfo {author} {\bibfnamefont {M.}~\bibnamefont
  {Hofheinz}},\ }\href {\doibase 10.1103/PhysRevX.9.021016} {\bibfield
  {journal} {\bibinfo  {journal} {Phys. Rev. X}\ }\textbf {\bibinfo {volume}
  {9}},\ \bibinfo {pages} {021016} (\bibinfo {year} {2019})}\BibitemShut
  {NoStop}%
\bibitem [{\citenamefont {Rolland}\ \emph {et~al.}(2019)\citenamefont
  {Rolland}, \citenamefont {Peugeot}, \citenamefont {Dambach}, \citenamefont
  {Westig}, \citenamefont {Kubala}, \citenamefont {Mukharsky}, \citenamefont
  {Altimiras}, \citenamefont {Le~Sueur}, \citenamefont {Joyez}, \citenamefont
  {Vion}, \citenamefont {Roche}, \citenamefont {Esteve}, \citenamefont
  {Ankerhold},\ and\ \citenamefont {Portier}}]{Rolland2019}%
  \BibitemOpen
  \bibfield  {author} {\bibinfo {author} {\bibfnamefont {C.}~\bibnamefont
  {Rolland}}, \bibinfo {author} {\bibfnamefont {A.}~\bibnamefont {Peugeot}},
  \bibinfo {author} {\bibfnamefont {S.}~\bibnamefont {Dambach}}, \bibinfo
  {author} {\bibfnamefont {M.}~\bibnamefont {Westig}}, \bibinfo {author}
  {\bibfnamefont {B.}~\bibnamefont {Kubala}}, \bibinfo {author} {\bibfnamefont
  {Y.}~\bibnamefont {Mukharsky}}, \bibinfo {author} {\bibfnamefont
  {C.}~\bibnamefont {Altimiras}}, \bibinfo {author} {\bibfnamefont
  {H.}~\bibnamefont {Le~Sueur}}, \bibinfo {author} {\bibfnamefont
  {P.}~\bibnamefont {Joyez}}, \bibinfo {author} {\bibfnamefont
  {D.}~\bibnamefont {Vion}}, \bibinfo {author} {\bibfnamefont {P.}~\bibnamefont
  {Roche}}, \bibinfo {author} {\bibfnamefont {D.}~\bibnamefont {Esteve}},
  \bibinfo {author} {\bibfnamefont {J.}~\bibnamefont {Ankerhold}}, \ and\
  \bibinfo {author} {\bibfnamefont {F.}~\bibnamefont {Portier}},\ }\href
  {\doibase https://doi.org/10.1103/PhysRevLett.122.186804} {\bibfield
  {journal} {\bibinfo  {journal} {Phys. Rev. Lett.}\ }\textbf {\bibinfo
  {volume} {122}},\ \bibinfo {pages} {186804} (\bibinfo {year}
  {2019})}\BibitemShut {NoStop}%
\bibitem [{\citenamefont {Westig}\ \emph {et~al.}(2017)\citenamefont {Westig},
  \citenamefont {Kubala}, \citenamefont {Parlavecchio}, \citenamefont
  {Mukharsky}, \citenamefont {Altimiras}, \citenamefont {Joyez}, \citenamefont
  {Vion}, \citenamefont {Roche}, \citenamefont {Esteve}, \citenamefont
  {Hofheinz}, \citenamefont {Trif}, \citenamefont {Simon}, \citenamefont
  {Ankerhold},\ and\ \citenamefont {Portier}}]{Westig2017}%
  \BibitemOpen
  \bibfield  {author} {\bibinfo {author} {\bibfnamefont {M.}~\bibnamefont
  {Westig}}, \bibinfo {author} {\bibfnamefont {B.}~\bibnamefont {Kubala}},
  \bibinfo {author} {\bibfnamefont {O.}~\bibnamefont {Parlavecchio}}, \bibinfo
  {author} {\bibfnamefont {Y.}~\bibnamefont {Mukharsky}}, \bibinfo {author}
  {\bibfnamefont {C.}~\bibnamefont {Altimiras}}, \bibinfo {author}
  {\bibfnamefont {P.}~\bibnamefont {Joyez}}, \bibinfo {author} {\bibfnamefont
  {D.}~\bibnamefont {Vion}}, \bibinfo {author} {\bibfnamefont {P.}~\bibnamefont
  {Roche}}, \bibinfo {author} {\bibfnamefont {D.}~\bibnamefont {Esteve}},
  \bibinfo {author} {\bibfnamefont {M.}~\bibnamefont {Hofheinz}}, \bibinfo
  {author} {\bibfnamefont {M.}~\bibnamefont {Trif}}, \bibinfo {author}
  {\bibfnamefont {P.}~\bibnamefont {Simon}}, \bibinfo {author} {\bibfnamefont
  {J.}~\bibnamefont {Ankerhold}}, \ and\ \bibinfo {author} {\bibfnamefont
  {F.}~\bibnamefont {Portier}},\ }\href {\doibase
  10.1103/PhysRevLett.119.137001} {\bibfield  {journal} {\bibinfo  {journal}
  {Phys. Rev. Lett.}\ }\textbf {\bibinfo {volume} {119}},\ \bibinfo {pages}
  {137001} (\bibinfo {year} {2017})}\BibitemShut {NoStop}%
\bibitem [{\citenamefont {Peugeot}\ \emph {et~al.}(2020)\citenamefont
  {Peugeot}, \citenamefont {M{\'e}nard}, \citenamefont {Dambach}, \citenamefont
  {Westig}, \citenamefont {Kubala}, \citenamefont {Mukharsky}, \citenamefont
  {Altimiras}, \citenamefont {Joyez}, \citenamefont {Vion}, \citenamefont
  {Roche}, \citenamefont {Esteve}, \citenamefont {Milman}, \citenamefont
  {Lepp{\"a}kangas}, \citenamefont {Johansson}, \citenamefont {Hofheinz},
  \citenamefont {Ankerhold},\ and\ \citenamefont {Portier}}]{Peugeot2020}%
  \BibitemOpen
  \bibfield  {author} {\bibinfo {author} {\bibfnamefont {A.}~\bibnamefont
  {Peugeot}}, \bibinfo {author} {\bibfnamefont {G.}~\bibnamefont {M{\'e}nard}},
  \bibinfo {author} {\bibfnamefont {S.}~\bibnamefont {Dambach}}, \bibinfo
  {author} {\bibfnamefont {M.}~\bibnamefont {Westig}}, \bibinfo {author}
  {\bibfnamefont {B.}~\bibnamefont {Kubala}}, \bibinfo {author} {\bibfnamefont
  {Y.}~\bibnamefont {Mukharsky}}, \bibinfo {author} {\bibfnamefont
  {C.}~\bibnamefont {Altimiras}}, \bibinfo {author} {\bibfnamefont
  {P.}~\bibnamefont {Joyez}}, \bibinfo {author} {\bibfnamefont
  {D.}~\bibnamefont {Vion}}, \bibinfo {author} {\bibfnamefont {P.}~\bibnamefont
  {Roche}}, \bibinfo {author} {\bibfnamefont {D.}~\bibnamefont {Esteve}},
  \bibinfo {author} {\bibfnamefont {P.}~\bibnamefont {Milman}}, \bibinfo
  {author} {\bibfnamefont {J.}~\bibnamefont {Lepp{\"a}kangas}}, \bibinfo
  {author} {\bibfnamefont {G.}~\bibnamefont {Johansson}}, \bibinfo {author}
  {\bibfnamefont {M.}~\bibnamefont {Hofheinz}}, \bibinfo {author}
  {\bibfnamefont {J.}~\bibnamefont {Ankerhold}}, \ and\ \bibinfo {author}
  {\bibfnamefont {F.}~\bibnamefont {Portier}},\ }\href@noop {} {\enquote
  {\bibinfo {title} {Generating two continuous entangled microwave beams using
  a dc-biased josephson junction},}\ } (\bibinfo {year} {2020}),\ \Eprint
  {http://arxiv.org/abs/2010.03376} {arXiv:2010.03376} \BibitemShut {NoStop}%
\bibitem [{\citenamefont {Padurariu}\ \emph {et~al.}(2012)\citenamefont
  {Padurariu}, \citenamefont {Hassler},\ and\ \citenamefont
  {Nazarov}}]{Padurariu2012}%
  \BibitemOpen
  \bibfield  {author} {\bibinfo {author} {\bibfnamefont {C.}~\bibnamefont
  {Padurariu}}, \bibinfo {author} {\bibfnamefont {F.}~\bibnamefont {Hassler}},
  \ and\ \bibinfo {author} {\bibfnamefont {Y.~V.}\ \bibnamefont {Nazarov}},\
  }\href {\doibase 10.1103/PhysRevB.86.054514} {\bibfield  {journal} {\bibinfo
  {journal} {Phys. Rev. B}\ }\textbf {\bibinfo {volume} {86}},\ \bibinfo
  {pages} {054514} (\bibinfo {year} {2012})}\BibitemShut {NoStop}%
\bibitem [{\citenamefont {Gramich}\ \emph {et~al.}(2013)\citenamefont
  {Gramich}, \citenamefont {Kubala}, \citenamefont {Rohrer},\ and\
  \citenamefont {Ankerhold}}]{Gramich2013}%
  \BibitemOpen
  \bibfield  {author} {\bibinfo {author} {\bibfnamefont {V.}~\bibnamefont
  {Gramich}}, \bibinfo {author} {\bibfnamefont {B.}~\bibnamefont {Kubala}},
  \bibinfo {author} {\bibfnamefont {S.}~\bibnamefont {Rohrer}}, \ and\ \bibinfo
  {author} {\bibfnamefont {J.}~\bibnamefont {Ankerhold}},\ }\href {\doibase
  10.1103/PhysRevLett.111.247002} {\bibfield  {journal} {\bibinfo  {journal}
  {Phys. Rev. Lett.}\ }\textbf {\bibinfo {volume} {111}},\ \bibinfo {pages}
  {247002} (\bibinfo {year} {2013})}\BibitemShut {NoStop}%
\bibitem [{\citenamefont {Armour}\ \emph {et~al.}(2013)\citenamefont {Armour},
  \citenamefont {Blencowe}, \citenamefont {Brahimi},\ and\ \citenamefont
  {Rimberg}}]{Armour2013}%
  \BibitemOpen
  \bibfield  {author} {\bibinfo {author} {\bibfnamefont {A.~D.}\ \bibnamefont
  {Armour}}, \bibinfo {author} {\bibfnamefont {M.~P.}\ \bibnamefont
  {Blencowe}}, \bibinfo {author} {\bibfnamefont {E.}~\bibnamefont {Brahimi}}, \
  and\ \bibinfo {author} {\bibfnamefont {A.~J.}\ \bibnamefont {Rimberg}},\
  }\href {\doibase 10.1103/PhysRevLett.111.247001} {\bibfield  {journal}
  {\bibinfo  {journal} {Phys. Rev. Lett.}\ }\textbf {\bibinfo {volume} {111}},\
  \bibinfo {pages} {247001} (\bibinfo {year} {2013})}\BibitemShut {NoStop}%
\bibitem [{\citenamefont {Lepp\"akangas}\ \emph {et~al.}(2013)\citenamefont
  {Lepp\"akangas}, \citenamefont {Johansson}, \citenamefont {Marthaler},\ and\
  \citenamefont {Fogelstr\"om}}]{Juha2013}%
  \BibitemOpen
  \bibfield  {author} {\bibinfo {author} {\bibfnamefont {J.}~\bibnamefont
  {Lepp\"akangas}}, \bibinfo {author} {\bibfnamefont {G.}~\bibnamefont
  {Johansson}}, \bibinfo {author} {\bibfnamefont {M.}~\bibnamefont
  {Marthaler}}, \ and\ \bibinfo {author} {\bibfnamefont {M.}~\bibnamefont
  {Fogelstr\"om}},\ }\href {\doibase 10.1103/PhysRevLett.110.267004} {\bibfield
   {journal} {\bibinfo  {journal} {Phys. Rev. Lett.}\ }\textbf {\bibinfo
  {volume} {110}},\ \bibinfo {pages} {267004} (\bibinfo {year}
  {2013})}\BibitemShut {NoStop}%
\bibitem [{\citenamefont {Kubala}\ \emph {et~al.}(2015)\citenamefont {Kubala},
  \citenamefont {Gramich},\ and\ \citenamefont {Ankerhold}}]{Kubala2015}%
  \BibitemOpen
  \bibfield  {author} {\bibinfo {author} {\bibfnamefont {B.}~\bibnamefont
  {Kubala}}, \bibinfo {author} {\bibfnamefont {V.}~\bibnamefont {Gramich}}, \
  and\ \bibinfo {author} {\bibfnamefont {J.}~\bibnamefont {Ankerhold}},\ }\href
  {\doibase 10.1088/0031-8949/2015/T165/014029} {\bibfield  {journal} {\bibinfo
   {journal} {Phys. Scr.}\ }\textbf {\bibinfo {volume} {T165}},\ \bibinfo
  {pages} {014029} (\bibinfo {year} {2015})}\BibitemShut {NoStop}%
\bibitem [{\citenamefont {Armour}\ \emph {et~al.}(2015)\citenamefont {Armour},
  \citenamefont {Kubala},\ and\ \citenamefont {Ankerhold}}]{Armour2015}%
  \BibitemOpen
  \bibfield  {author} {\bibinfo {author} {\bibfnamefont {A.~D.}\ \bibnamefont
  {Armour}}, \bibinfo {author} {\bibfnamefont {B.}~\bibnamefont {Kubala}}, \
  and\ \bibinfo {author} {\bibfnamefont {J.}~\bibnamefont {Ankerhold}},\ }\href
  {\doibase 10.1103/PhysRevB.91.184508} {\bibfield  {journal} {\bibinfo
  {journal} {Phys. Rev. B}\ }\textbf {\bibinfo {volume} {91}},\ \bibinfo
  {pages} {184508} (\bibinfo {year} {2015})}\BibitemShut {NoStop}%
\bibitem [{\citenamefont {Trif}\ and\ \citenamefont {Simon}(2015)}]{Trif2015}%
  \BibitemOpen
  \bibfield  {author} {\bibinfo {author} {\bibfnamefont {M.}~\bibnamefont
  {Trif}}\ and\ \bibinfo {author} {\bibfnamefont {P.}~\bibnamefont {Simon}},\
  }\href {\doibase 10.1103/PhysRevB.92.014503} {\bibfield  {journal} {\bibinfo
  {journal} {Phys. Rev. B}\ }\textbf {\bibinfo {volume} {92}},\ \bibinfo
  {pages} {014503} (\bibinfo {year} {2015})}\BibitemShut {NoStop}%
\bibitem [{\citenamefont {Meister}\ \emph {et~al.}(2015)\citenamefont
  {Meister}, \citenamefont {Mecklenburg}, \citenamefont {Gramich},
  \citenamefont {Stockburger}, \citenamefont {Ankerhold},\ and\ \citenamefont
  {Kubala}}]{Meister2015}%
  \BibitemOpen
  \bibfield  {author} {\bibinfo {author} {\bibfnamefont {S.}~\bibnamefont
  {Meister}}, \bibinfo {author} {\bibfnamefont {M.}~\bibnamefont
  {Mecklenburg}}, \bibinfo {author} {\bibfnamefont {V.}~\bibnamefont
  {Gramich}}, \bibinfo {author} {\bibfnamefont {J.~T.}\ \bibnamefont
  {Stockburger}}, \bibinfo {author} {\bibfnamefont {J.}~\bibnamefont
  {Ankerhold}}, \ and\ \bibinfo {author} {\bibfnamefont {B.}~\bibnamefont
  {Kubala}},\ }\href {\doibase 10.1103/PhysRevB.92.174532} {\bibfield
  {journal} {\bibinfo  {journal} {Phys. Rev. B}\ }\textbf {\bibinfo {volume}
  {92}},\ \bibinfo {pages} {174532} (\bibinfo {year} {2015})}\BibitemShut
  {NoStop}%
\bibitem [{\citenamefont {Souquet}\ and\ \citenamefont
  {Clerk}(2016)}]{Souquet2016}%
  \BibitemOpen
  \bibfield  {author} {\bibinfo {author} {\bibfnamefont {J.-R.}\ \bibnamefont
  {Souquet}}\ and\ \bibinfo {author} {\bibfnamefont {A.~A.}\ \bibnamefont
  {Clerk}},\ }\href {\doibase 10.1103/PhysRevA.93.060301} {\bibfield  {journal}
  {\bibinfo  {journal} {Phys. Rev. A}\ }\textbf {\bibinfo {volume} {93}},\
  \bibinfo {pages} {060301} (\bibinfo {year} {2016})}\BibitemShut {NoStop}%
\bibitem [{\citenamefont {Lepp\"akangas}\ \emph {et~al.}(2016)\citenamefont
  {Lepp\"akangas}, \citenamefont {Fogelstr\"om}, \citenamefont {Marthaler},\
  and\ \citenamefont {Johansson}}]{Juha2016}%
  \BibitemOpen
  \bibfield  {author} {\bibinfo {author} {\bibfnamefont {J.}~\bibnamefont
  {Lepp\"akangas}}, \bibinfo {author} {\bibfnamefont {M.}~\bibnamefont
  {Fogelstr\"om}}, \bibinfo {author} {\bibfnamefont {M.}~\bibnamefont
  {Marthaler}}, \ and\ \bibinfo {author} {\bibfnamefont {G.}~\bibnamefont
  {Johansson}},\ }\href {\doibase 10.1103/PhysRevB.93.014506} {\bibfield
  {journal} {\bibinfo  {journal} {Phys. Rev. B}\ }\textbf {\bibinfo {volume}
  {93}},\ \bibinfo {pages} {014506} (\bibinfo {year} {2016})}\BibitemShut
  {NoStop}%
\bibitem [{\citenamefont {Armour}\ \emph {et~al.}(2017)\citenamefont {Armour},
  \citenamefont {Kubala},\ and\ \citenamefont {Ankerhold}}]{Armour2017}%
  \BibitemOpen
  \bibfield  {author} {\bibinfo {author} {\bibfnamefont {A.~D.}\ \bibnamefont
  {Armour}}, \bibinfo {author} {\bibfnamefont {B.}~\bibnamefont {Kubala}}, \
  and\ \bibinfo {author} {\bibfnamefont {J.}~\bibnamefont {Ankerhold}},\ }\href
  {\doibase 10.1103/PhysRevB.96.214509} {\bibfield  {journal} {\bibinfo
  {journal} {Phys. Rev. B}\ }\textbf {\bibinfo {volume} {96}},\ \bibinfo
  {pages} {214509} (\bibinfo {year} {2017})}\BibitemShut {NoStop}%
\bibitem [{\citenamefont {Wang}\ \emph {et~al.}(2017)\citenamefont {Wang},
  \citenamefont {Blencowe}, \citenamefont {Armour},\ and\ \citenamefont
  {Rimberg}}]{Wang2017}%
  \BibitemOpen
  \bibfield  {author} {\bibinfo {author} {\bibfnamefont {H.}~\bibnamefont
  {Wang}}, \bibinfo {author} {\bibfnamefont {M.~P.}\ \bibnamefont {Blencowe}},
  \bibinfo {author} {\bibfnamefont {A.~D.}\ \bibnamefont {Armour}}, \ and\
  \bibinfo {author} {\bibfnamefont {A.~J.}\ \bibnamefont {Rimberg}},\ }\href
  {\doibase 10.1103/PhysRevB.96.104503} {\bibfield  {journal} {\bibinfo
  {journal} {Phys. Rev. B}\ }\textbf {\bibinfo {volume} {96}},\ \bibinfo
  {pages} {104503} (\bibinfo {year} {2017})}\BibitemShut {NoStop}%
\bibitem [{\citenamefont {Simon}\ and\ \citenamefont
  {Cooper}(2018)}]{Simon2018}%
  \BibitemOpen
  \bibfield  {author} {\bibinfo {author} {\bibfnamefont {S.~H.}\ \bibnamefont
  {Simon}}\ and\ \bibinfo {author} {\bibfnamefont {N.~R.}\ \bibnamefont
  {Cooper}},\ }\href {\doibase 10.1103/PhysRevLett.121.027004} {\bibfield
  {journal} {\bibinfo  {journal} {Phys. Rev. Lett.}\ }\textbf {\bibinfo
  {volume} {121}},\ \bibinfo {pages} {027004} (\bibinfo {year}
  {2018})}\BibitemShut {NoStop}%
\bibitem [{\citenamefont {Lepp\"akangas}\ \emph {et~al.}(2018)\citenamefont
  {Lepp\"akangas}, \citenamefont {Marthaler}, \citenamefont {Hazra},
  \citenamefont {Jebari}, \citenamefont {Albert}, \citenamefont {Blanchet},
  \citenamefont {Johansson},\ and\ \citenamefont {Hofheinz}}]{Juha2018}%
  \BibitemOpen
  \bibfield  {author} {\bibinfo {author} {\bibfnamefont {J.}~\bibnamefont
  {Lepp\"akangas}}, \bibinfo {author} {\bibfnamefont {M.}~\bibnamefont
  {Marthaler}}, \bibinfo {author} {\bibfnamefont {D.}~\bibnamefont {Hazra}},
  \bibinfo {author} {\bibfnamefont {S.}~\bibnamefont {Jebari}}, \bibinfo
  {author} {\bibfnamefont {R.}~\bibnamefont {Albert}}, \bibinfo {author}
  {\bibfnamefont {F.}~\bibnamefont {Blanchet}}, \bibinfo {author}
  {\bibfnamefont {G.}~\bibnamefont {Johansson}}, \ and\ \bibinfo {author}
  {\bibfnamefont {M.}~\bibnamefont {Hofheinz}},\ }\href {\doibase
  10.1103/PhysRevA.97.013855} {\bibfield  {journal} {\bibinfo  {journal} {Phys.
  Rev. A}\ }\textbf {\bibinfo {volume} {97}},\ \bibinfo {pages} {013855}
  (\bibinfo {year} {2018})}\BibitemShut {NoStop}%
\bibitem [{\citenamefont {Arndt}\ and\ \citenamefont
  {Hassler}(2019)}]{Arndt2019}%
  \BibitemOpen
  \bibfield  {author} {\bibinfo {author} {\bibfnamefont {L.}~\bibnamefont
  {Arndt}}\ and\ \bibinfo {author} {\bibfnamefont {F.}~\bibnamefont
  {Hassler}},\ }\href {\doibase 10.1103/PhysRevB.100.014505} {\bibfield
  {journal} {\bibinfo  {journal} {Phys. Rev. B}\ }\textbf {\bibinfo {volume}
  {100}},\ \bibinfo {pages} {014505} (\bibinfo {year} {2019})}\BibitemShut
  {NoStop}%
\bibitem [{\citenamefont {Morley}\ \emph {et~al.}(2019)\citenamefont {Morley},
  \citenamefont {Di~Marco}, \citenamefont {Mantovani}, \citenamefont {Stadler},
  \citenamefont {Belzig}, \citenamefont {Rastelli},\ and\ \citenamefont
  {Armour}}]{Morley2019}%
  \BibitemOpen
  \bibfield  {author} {\bibinfo {author} {\bibfnamefont {W.~T.}\ \bibnamefont
  {Morley}}, \bibinfo {author} {\bibfnamefont {A.}~\bibnamefont {Di~Marco}},
  \bibinfo {author} {\bibfnamefont {M.}~\bibnamefont {Mantovani}}, \bibinfo
  {author} {\bibfnamefont {P.}~\bibnamefont {Stadler}}, \bibinfo {author}
  {\bibfnamefont {W.}~\bibnamefont {Belzig}}, \bibinfo {author} {\bibfnamefont
  {G.}~\bibnamefont {Rastelli}}, \ and\ \bibinfo {author} {\bibfnamefont
  {A.~D.}\ \bibnamefont {Armour}},\ }\href {\doibase
  10.1103/PhysRevB.100.054515} {\bibfield  {journal} {\bibinfo  {journal}
  {Phys. Rev. B}\ }\textbf {\bibinfo {volume} {100}},\ \bibinfo {pages}
  {054515} (\bibinfo {year} {2019})}\BibitemShut {NoStop}%
\bibitem [{\citenamefont {Kubala}\ \emph {et~al.}(2020)\citenamefont {Kubala},
  \citenamefont {Ankerhold},\ and\ \citenamefont {Armour}}]{Kubala2020}%
  \BibitemOpen
  \bibfield  {author} {\bibinfo {author} {\bibfnamefont {B.}~\bibnamefont
  {Kubala}}, \bibinfo {author} {\bibfnamefont {J.}~\bibnamefont {Ankerhold}}, \
  and\ \bibinfo {author} {\bibfnamefont {A.~D.}\ \bibnamefont {Armour}},\
  }\href {\doibase 10.1088/1367-2630/ab6eb0} {\bibfield  {journal} {\bibinfo
  {journal} {New Journal of Physics}\ }\textbf {\bibinfo {volume} {22}},\
  \bibinfo {pages} {023010} (\bibinfo {year} {2020})}\BibitemShut {NoStop}%
\bibitem [{\citenamefont {Lang}\ and\ \citenamefont {Armour}(2020)}]{Lang2020}%
  \BibitemOpen
  \bibfield  {author} {\bibinfo {author} {\bibfnamefont {B.}~\bibnamefont
  {Lang}}\ and\ \bibinfo {author} {\bibfnamefont {A.~D.}\ \bibnamefont
  {Armour}},\ }\href@noop {} {\enquote {\bibinfo {title} {Multi-photon
  resonances in josephson junction-cavity circuits},}\ } (\bibinfo {year}
  {2020}),\ \Eprint {http://arxiv.org/abs/2012.10149} {arXiv:2012.10149
  [cond-mat.mes-hall]} \BibitemShut {NoStop}%
\bibitem [{\citenamefont {Hriscu}\ and\ \citenamefont
  {Nazarov}(2013)}]{Hriscu2013}%
  \BibitemOpen
  \bibfield  {author} {\bibinfo {author} {\bibfnamefont {A.~M.}\ \bibnamefont
  {Hriscu}}\ and\ \bibinfo {author} {\bibfnamefont {Y.~V.}\ \bibnamefont
  {Nazarov}},\ }\href {\doibase 10.1103/PhysRevLett.110.097002} {\bibfield
  {journal} {\bibinfo  {journal} {Phys. Rev. Lett.}\ }\textbf {\bibinfo
  {volume} {110}},\ \bibinfo {pages} {097002} (\bibinfo {year}
  {2013})}\BibitemShut {NoStop}%
\bibitem [{\citenamefont {Bezryadin}\ \emph {et~al.}(2000)\citenamefont
  {Bezryadin}, \citenamefont {Lau},\ and\ \citenamefont
  {Tinkham}}]{Bezryadin2000}%
  \BibitemOpen
  \bibfield  {author} {\bibinfo {author} {\bibfnamefont {A.}~\bibnamefont
  {Bezryadin}}, \bibinfo {author} {\bibfnamefont {C.~N.}\ \bibnamefont {Lau}},
  \ and\ \bibinfo {author} {\bibfnamefont {M.}~\bibnamefont {Tinkham}},\ }\href
  {\doibase 10.1038/35010060} {\bibfield  {journal} {\bibinfo  {journal}
  {Nature}\ }\textbf {\bibinfo {volume} {404}},\ \bibinfo {pages} {971}
  (\bibinfo {year} {2000})}\BibitemShut {NoStop}%
\bibitem [{\citenamefont {Altomare}\ \emph {et~al.}(2006)\citenamefont
  {Altomare}, \citenamefont {Chang}, \citenamefont {Melloch}, \citenamefont
  {Hong},\ and\ \citenamefont {Tu}}]{Altomare2006}%
  \BibitemOpen
  \bibfield  {author} {\bibinfo {author} {\bibfnamefont {F.}~\bibnamefont
  {Altomare}}, \bibinfo {author} {\bibfnamefont {A.~M.}\ \bibnamefont {Chang}},
  \bibinfo {author} {\bibfnamefont {M.~R.}\ \bibnamefont {Melloch}}, \bibinfo
  {author} {\bibfnamefont {Y.}~\bibnamefont {Hong}}, \ and\ \bibinfo {author}
  {\bibfnamefont {C.~W.}\ \bibnamefont {Tu}},\ }\href {\doibase
  10.1103/PhysRevLett.97.017001} {\bibfield  {journal} {\bibinfo  {journal}
  {Phys. Rev. Lett.}\ }\textbf {\bibinfo {volume} {97}},\ \bibinfo {pages}
  {017001} (\bibinfo {year} {2006})}\BibitemShut {NoStop}%
\bibitem [{\citenamefont {Astafiev}\ \emph {et~al.}(2012)\citenamefont
  {Astafiev}, \citenamefont {Ioffe}, \citenamefont {Kafanov}, \citenamefont
  {Pashkin}, \citenamefont {Arutyunov}, \citenamefont {Shahar}, \citenamefont
  {Cohen},\ and\ \citenamefont {Tsai}}]{Astafiev2012}%
  \BibitemOpen
  \bibfield  {author} {\bibinfo {author} {\bibfnamefont {O.~V.}\ \bibnamefont
  {Astafiev}}, \bibinfo {author} {\bibfnamefont {L.~B.}\ \bibnamefont {Ioffe}},
  \bibinfo {author} {\bibfnamefont {S.}~\bibnamefont {Kafanov}}, \bibinfo
  {author} {\bibfnamefont {Y.~A.}\ \bibnamefont {Pashkin}}, \bibinfo {author}
  {\bibfnamefont {K.~Y.}\ \bibnamefont {Arutyunov}}, \bibinfo {author}
  {\bibfnamefont {D.}~\bibnamefont {Shahar}}, \bibinfo {author} {\bibfnamefont
  {O.}~\bibnamefont {Cohen}}, \ and\ \bibinfo {author} {\bibfnamefont {J.~S.}\
  \bibnamefont {Tsai}},\ }\href {\doibase 10.1038/nature10930} {\bibfield
  {journal} {\bibinfo  {journal} {Nature}\ }\textbf {\bibinfo {volume} {484}},\
  \bibinfo {pages} {355} (\bibinfo {year} {2012})}\BibitemShut {NoStop}%
\bibitem [{\citenamefont {Chen}\ \emph
  {et~al.}(2014{\natexlab{b}})\citenamefont {Chen}, \citenamefont {Lin},
  \citenamefont {Snyder}, \citenamefont {Goldman},\ and\ \citenamefont
  {Kamenev}}]{Chen2014b}%
  \BibitemOpen
  \bibfield  {author} {\bibinfo {author} {\bibfnamefont {Y.}~\bibnamefont
  {Chen}}, \bibinfo {author} {\bibfnamefont {Y.-H.}\ \bibnamefont {Lin}},
  \bibinfo {author} {\bibfnamefont {S.~D.}\ \bibnamefont {Snyder}}, \bibinfo
  {author} {\bibfnamefont {A.~M.}\ \bibnamefont {Goldman}}, \ and\ \bibinfo
  {author} {\bibfnamefont {A.}~\bibnamefont {Kamenev}},\ }\href {\doibase
  10.1038/nphys3008} {\bibfield  {journal} {\bibinfo  {journal} {Nature
  Physics}\ }\textbf {\bibinfo {volume} {10}},\ \bibinfo {pages} {567}
  (\bibinfo {year} {2014}{\natexlab{b}})}\BibitemShut {NoStop}%
\bibitem [{\citenamefont {{Adler}}(1946)}]{Adler1964}%
  \BibitemOpen
  \bibfield  {author} {\bibinfo {author} {\bibfnamefont {R.}~\bibnamefont
  {{Adler}}},\ }\href@noop {} {\bibfield  {journal} {\bibinfo  {journal}
  {Proceedings of the IRE}\ }\textbf {\bibinfo {volume} {34}},\ \bibinfo
  {pages} {351} (\bibinfo {year} {1946})}\BibitemShut {NoStop}%
\bibitem [{\citenamefont {Pikovsky}\ \emph {et~al.}(2001)\citenamefont
  {Pikovsky}, \citenamefont {Rosenblum},\ and\ \citenamefont
  {Kurths}}]{Pikovsky2001}%
  \BibitemOpen
  \bibfield  {author} {\bibinfo {author} {\bibfnamefont {A.}~\bibnamefont
  {Pikovsky}}, \bibinfo {author} {\bibfnamefont {M.~G.}\ \bibnamefont
  {Rosenblum}}, \ and\ \bibinfo {author} {\bibfnamefont {J.}~\bibnamefont
  {Kurths}},\ }\href@noop {} {\emph {\bibinfo {title} {Synchronization, A
  Universal Concept in Nonlinear Sciences}}}\ (\bibinfo  {publisher} {Cambridge
  University Press},\ \bibinfo {address} {Cambridge},\ \bibinfo {year}
  {2001})\BibitemShut {NoStop}%
\bibitem [{Note1()}]{Note1}%
  \BibitemOpen
  \bibinfo {note} {It has been noted, however, that a model without $R_0$ and a
  strictly fixed dc-bias would result in emission without spectral width, and
  the observed spectral linewidth, typically much sharper than the inverse
  cavity lifetime, has been associated with low-frequency fluctuations of the
  voltage at the junction}\BibitemShut {NoStop}%
\bibitem [{Note2()}]{Note2}%
  \BibitemOpen
  \bibinfo {note} {Note, that there is a subtlety in defining the phase of the
  rotating frame and there is also a valid approach which results in a limit
  cycle description for the $R_0=0$ case, but does not lead to
  locking.}\BibitemShut {Stop}%
\bibitem [{\citenamefont {Shapiro}(1963)}]{Shapiro1963}%
  \BibitemOpen
  \bibfield  {author} {\bibinfo {author} {\bibfnamefont {S.}~\bibnamefont
  {Shapiro}},\ }\href {\doibase 10.1103/PhysRevLett.11.80} {\bibfield
  {journal} {\bibinfo  {journal} {Phys. Rev. Lett.}\ }\textbf {\bibinfo
  {volume} {11}},\ \bibinfo {pages} {80} (\bibinfo {year} {1963})}\BibitemShut
  {NoStop}%
\bibitem [{\citenamefont {Kramers}(1940)}]{Kramers1940}%
  \BibitemOpen
  \bibfield  {author} {\bibinfo {author} {\bibfnamefont {H.~A.}\ \bibnamefont
  {Kramers}},\ }\href@noop {} {\bibfield  {journal} {\bibinfo  {journal}
  {Physica}\ }\textbf {\bibinfo {volume} {7}},\ \bibinfo {pages} {284}
  (\bibinfo {year} {1940})}\BibitemShut {NoStop}%
\bibitem [{\citenamefont {Walter}\ \emph {et~al.}(2014)\citenamefont {Walter},
  \citenamefont {Nunnenkamp},\ and\ \citenamefont {Bruder}}]{Walter2014}%
  \BibitemOpen
  \bibfield  {author} {\bibinfo {author} {\bibfnamefont {S.}~\bibnamefont
  {Walter}}, \bibinfo {author} {\bibfnamefont {A.}~\bibnamefont {Nunnenkamp}},
  \ and\ \bibinfo {author} {\bibfnamefont {C.}~\bibnamefont {Bruder}},\ }\href
  {\doibase 10.1103/PhysRevLett.112.094102} {\bibfield  {journal} {\bibinfo
  {journal} {Phys. Rev. Lett.}\ }\textbf {\bibinfo {volume} {112}},\ \bibinfo
  {pages} {094102} (\bibinfo {year} {2014})}\BibitemShut {NoStop}%
\bibitem [{\citenamefont {L\"orch}\ \emph {et~al.}(2016)\citenamefont
  {L\"orch}, \citenamefont {Amitai}, \citenamefont {Nunnenkamp},\ and\
  \citenamefont {Bruder}}]{Loerch2016}%
  \BibitemOpen
  \bibfield  {author} {\bibinfo {author} {\bibfnamefont {N.}~\bibnamefont
  {L\"orch}}, \bibinfo {author} {\bibfnamefont {E.}~\bibnamefont {Amitai}},
  \bibinfo {author} {\bibfnamefont {A.}~\bibnamefont {Nunnenkamp}}, \ and\
  \bibinfo {author} {\bibfnamefont {C.}~\bibnamefont {Bruder}},\ }\href
  {\doibase 10.1103/PhysRevLett.117.073601} {\bibfield  {journal} {\bibinfo
  {journal} {Phys. Rev. Lett.}\ }\textbf {\bibinfo {volume} {117}},\ \bibinfo
  {pages} {073601} (\bibinfo {year} {2016})}\BibitemShut {NoStop}%
\bibitem [{\citenamefont {L\"orch}\ \emph {et~al.}(2017)\citenamefont
  {L\"orch}, \citenamefont {Nigg}, \citenamefont {Nunnenkamp}, \citenamefont
  {Tiwari},\ and\ \citenamefont {Bruder}}]{Loerch2017}%
  \BibitemOpen
  \bibfield  {author} {\bibinfo {author} {\bibfnamefont {N.}~\bibnamefont
  {L\"orch}}, \bibinfo {author} {\bibfnamefont {S.~E.}\ \bibnamefont {Nigg}},
  \bibinfo {author} {\bibfnamefont {A.}~\bibnamefont {Nunnenkamp}}, \bibinfo
  {author} {\bibfnamefont {R.~P.}\ \bibnamefont {Tiwari}}, \ and\ \bibinfo
  {author} {\bibfnamefont {C.}~\bibnamefont {Bruder}},\ }\href {\doibase
  10.1103/PhysRevLett.118.243602} {\bibfield  {journal} {\bibinfo  {journal}
  {Phys. Rev. Lett.}\ }\textbf {\bibinfo {volume} {118}},\ \bibinfo {pages}
  {243602} (\bibinfo {year} {2017})}\BibitemShut {NoStop}%
\bibitem [{\citenamefont {Roulet}\ and\ \citenamefont
  {Bruder}(2018)}]{Roulet2018}%
  \BibitemOpen
  \bibfield  {author} {\bibinfo {author} {\bibfnamefont {A.}~\bibnamefont
  {Roulet}}\ and\ \bibinfo {author} {\bibfnamefont {C.}~\bibnamefont
  {Bruder}},\ }\href {\doibase 10.1103/PhysRevLett.121.053601} {\bibfield
  {journal} {\bibinfo  {journal} {Phys. Rev. Lett.}\ }\textbf {\bibinfo
  {volume} {121}},\ \bibinfo {pages} {053601} (\bibinfo {year}
  {2018})}\BibitemShut {NoStop}%
\bibitem [{\citenamefont {Amitai}\ \emph {et~al.}(2018)\citenamefont {Amitai},
  \citenamefont {Koppenh\"ofer}, \citenamefont {L\"orch},\ and\ \citenamefont
  {Bruder}}]{Amitai2018}%
  \BibitemOpen
  \bibfield  {author} {\bibinfo {author} {\bibfnamefont {E.}~\bibnamefont
  {Amitai}}, \bibinfo {author} {\bibfnamefont {M.}~\bibnamefont
  {Koppenh\"ofer}}, \bibinfo {author} {\bibfnamefont {N.}~\bibnamefont
  {L\"orch}}, \ and\ \bibinfo {author} {\bibfnamefont {C.}~\bibnamefont
  {Bruder}},\ }\href {\doibase 10.1103/PhysRevE.97.052203} {\bibfield
  {journal} {\bibinfo  {journal} {Phys. Rev. E}\ }\textbf {\bibinfo {volume}
  {97}},\ \bibinfo {pages} {052203} (\bibinfo {year} {2018})}\BibitemShut
  {NoStop}%
\bibitem [{\citenamefont {Koppenh\"ofer}\ and\ \citenamefont
  {Roulet}(2019)}]{Koppenhoefer2019}%
  \BibitemOpen
  \bibfield  {author} {\bibinfo {author} {\bibfnamefont {M.}~\bibnamefont
  {Koppenh\"ofer}}\ and\ \bibinfo {author} {\bibfnamefont {A.}~\bibnamefont
  {Roulet}},\ }\href {\doibase 10.1103/PhysRevA.99.043804} {\bibfield
  {journal} {\bibinfo  {journal} {Phys. Rev. A}\ }\textbf {\bibinfo {volume}
  {99}},\ \bibinfo {pages} {043804} (\bibinfo {year} {2019})}\BibitemShut
  {NoStop}%
\bibitem [{\citenamefont {Jessop}\ \emph {et~al.}(2020)\citenamefont {Jessop},
  \citenamefont {Li},\ and\ \citenamefont {Armour}}]{Jessop2020}%
  \BibitemOpen
  \bibfield  {author} {\bibinfo {author} {\bibfnamefont {M.~R.}\ \bibnamefont
  {Jessop}}, \bibinfo {author} {\bibfnamefont {W.}~\bibnamefont {Li}}, \ and\
  \bibinfo {author} {\bibfnamefont {A.~D.}\ \bibnamefont {Armour}},\ }\href
  {\doibase 10.1103/PhysRevResearch.2.013233} {\bibfield  {journal} {\bibinfo
  {journal} {Phys. Rev. Research}\ }\textbf {\bibinfo {volume} {2}},\ \bibinfo
  {pages} {013233} (\bibinfo {year} {2020})}\BibitemShut {NoStop}%
\bibitem [{\citenamefont {Ben~Arosh}\ \emph {et~al.}(2021)\citenamefont
  {Ben~Arosh}, \citenamefont {Cross},\ and\ \citenamefont
  {Lifshitz}}]{Lifshitz2021}%
  \BibitemOpen
  \bibfield  {author} {\bibinfo {author} {\bibfnamefont {L.}~\bibnamefont
  {Ben~Arosh}}, \bibinfo {author} {\bibfnamefont {M.~C.}\ \bibnamefont
  {Cross}}, \ and\ \bibinfo {author} {\bibfnamefont {R.}~\bibnamefont
  {Lifshitz}},\ }\href {\doibase 10.1103/PhysRevResearch.3.013130} {\bibfield
  {journal} {\bibinfo  {journal} {Phys. Rev. Research}\ }\textbf {\bibinfo
  {volume} {3}},\ \bibinfo {pages} {013130} (\bibinfo {year}
  {2021})}\BibitemShut {NoStop}%
\bibitem [{\citenamefont {Wustmann}\ and\ \citenamefont
  {Shumeiko}(2019)}]{Wustmann2019}%
  \BibitemOpen
  \bibfield  {author} {\bibinfo {author} {\bibfnamefont {W.}~\bibnamefont
  {Wustmann}}\ and\ \bibinfo {author} {\bibfnamefont {V.}~\bibnamefont
  {Shumeiko}},\ }\href {\doibase 10.1063/1.5116533} {\bibfield  {journal}
  {\bibinfo  {journal} {Low Temperature Physics}\ }\textbf {\bibinfo {volume}
  {45}},\ \bibinfo {pages} {848} (\bibinfo {year} {2019})},\ \Eprint
  {http://arxiv.org/abs/https://doi.org/10.1063/1.5116533}
  {https://doi.org/10.1063/1.5116533} \BibitemShut {NoStop}%
\bibitem [{\citenamefont {Svensson}\ \emph {et~al.}(2017)\citenamefont
  {Svensson}, \citenamefont {Bengtsson}, \citenamefont {Krantz}, \citenamefont
  {Bylander}, \citenamefont {Shumeiko},\ and\ \citenamefont
  {Delsing}}]{Svenson2017}%
  \BibitemOpen
  \bibfield  {author} {\bibinfo {author} {\bibfnamefont {I.-M.}\ \bibnamefont
  {Svensson}}, \bibinfo {author} {\bibfnamefont {A.}~\bibnamefont {Bengtsson}},
  \bibinfo {author} {\bibfnamefont {P.}~\bibnamefont {Krantz}}, \bibinfo
  {author} {\bibfnamefont {J.}~\bibnamefont {Bylander}}, \bibinfo {author}
  {\bibfnamefont {V.}~\bibnamefont {Shumeiko}}, \ and\ \bibinfo {author}
  {\bibfnamefont {P.}~\bibnamefont {Delsing}},\ }\href {\doibase
  10.1103/PhysRevB.96.174503} {\bibfield  {journal} {\bibinfo  {journal} {Phys.
  Rev. B}\ }\textbf {\bibinfo {volume} {96}},\ \bibinfo {pages} {174503}
  (\bibinfo {year} {2017})}\BibitemShut {NoStop}%
\bibitem [{\citenamefont {Svensson}\ \emph {et~al.}(2018)\citenamefont
  {Svensson}, \citenamefont {Bengtsson}, \citenamefont {Bylander},
  \citenamefont {Shumeiko},\ and\ \citenamefont {Delsing}}]{Svenson2018}%
  \BibitemOpen
  \bibfield  {author} {\bibinfo {author} {\bibfnamefont {I.-M.}\ \bibnamefont
  {Svensson}}, \bibinfo {author} {\bibfnamefont {A.}~\bibnamefont {Bengtsson}},
  \bibinfo {author} {\bibfnamefont {J.}~\bibnamefont {Bylander}}, \bibinfo
  {author} {\bibfnamefont {V.}~\bibnamefont {Shumeiko}}, \ and\ \bibinfo
  {author} {\bibfnamefont {P.}~\bibnamefont {Delsing}},\ }\href {\doibase
  10.1063/1.5026974} {\bibfield  {journal} {\bibinfo  {journal} {Applied
  Physics Letters}\ }\textbf {\bibinfo {volume} {113}},\ \bibinfo {pages}
  {022602} (\bibinfo {year} {2018})},\ \Eprint
  {http://arxiv.org/abs/https://doi.org/10.1063/1.5026974}
  {https://doi.org/10.1063/1.5026974} \BibitemShut {NoStop}%
\bibitem [{\citenamefont {Chang}\ \emph {et~al.}(2020)\citenamefont {Chang},
  \citenamefont {Sab\'{\i}n}, \citenamefont {Forn-D\'{\i}az}, \citenamefont
  {Quijandr\'{\i}a}, \citenamefont {Vadiraj}, \citenamefont {Nsanzineza},
  \citenamefont {Johansson},\ and\ \citenamefont {Wilson}}]{Chang2020}%
  \BibitemOpen
  \bibfield  {author} {\bibinfo {author} {\bibfnamefont {C.~W.~S.}\
  \bibnamefont {Chang}}, \bibinfo {author} {\bibfnamefont {C.}~\bibnamefont
  {Sab\'{\i}n}}, \bibinfo {author} {\bibfnamefont {P.}~\bibnamefont
  {Forn-D\'{\i}az}}, \bibinfo {author} {\bibfnamefont {F.}~\bibnamefont
  {Quijandr\'{\i}a}}, \bibinfo {author} {\bibfnamefont {A.~M.}\ \bibnamefont
  {Vadiraj}}, \bibinfo {author} {\bibfnamefont {I.}~\bibnamefont {Nsanzineza}},
  \bibinfo {author} {\bibfnamefont {G.}~\bibnamefont {Johansson}}, \ and\
  \bibinfo {author} {\bibfnamefont {C.~M.}\ \bibnamefont {Wilson}},\ }\href
  {\doibase 10.1103/PhysRevX.10.011011} {\bibfield  {journal} {\bibinfo
  {journal} {Phys. Rev. X}\ }\textbf {\bibinfo {volume} {10}},\ \bibinfo
  {pages} {011011} (\bibinfo {year} {2020})}\BibitemShut {NoStop}%
\bibitem [{\citenamefont {Guo}\ \emph {et~al.}(2013)\citenamefont {Guo},
  \citenamefont {Marthaler},\ and\ \citenamefont {Sch\"on}}]{Guo2013}%
  \BibitemOpen
  \bibfield  {author} {\bibinfo {author} {\bibfnamefont {L.}~\bibnamefont
  {Guo}}, \bibinfo {author} {\bibfnamefont {M.}~\bibnamefont {Marthaler}}, \
  and\ \bibinfo {author} {\bibfnamefont {G.}~\bibnamefont {Sch\"on}},\ }\href
  {\doibase 10.1103/PhysRevLett.111.205303} {\bibfield  {journal} {\bibinfo
  {journal} {Phys. Rev. Lett.}\ }\textbf {\bibinfo {volume} {111}},\ \bibinfo
  {pages} {205303} (\bibinfo {year} {2013})}\BibitemShut {NoStop}%
\bibitem [{\citenamefont {Zhang}\ \emph {et~al.}(2017)\citenamefont {Zhang},
  \citenamefont {Gosner}, \citenamefont {Girvin}, \citenamefont {Ankerhold},\
  and\ \citenamefont {Dykman}}]{Zhang2017}%
  \BibitemOpen
  \bibfield  {author} {\bibinfo {author} {\bibfnamefont {Y.}~\bibnamefont
  {Zhang}}, \bibinfo {author} {\bibfnamefont {J.}~\bibnamefont {Gosner}},
  \bibinfo {author} {\bibfnamefont {S.~M.}\ \bibnamefont {Girvin}}, \bibinfo
  {author} {\bibfnamefont {J.}~\bibnamefont {Ankerhold}}, \ and\ \bibinfo
  {author} {\bibfnamefont {M.~I.}\ \bibnamefont {Dykman}},\ }\href {\doibase
  10.1103/PhysRevA.96.052124} {\bibfield  {journal} {\bibinfo  {journal} {Phys.
  Rev. A}\ }\textbf {\bibinfo {volume} {96}},\ \bibinfo {pages} {052124}
  (\bibinfo {year} {2017})}\BibitemShut {NoStop}%
\bibitem [{\citenamefont {Gosner}\ \emph {et~al.}(2020)\citenamefont {Gosner},
  \citenamefont {Kubala},\ and\ \citenamefont {Ankerhold}}]{Gosner2020}%
  \BibitemOpen
  \bibfield  {author} {\bibinfo {author} {\bibfnamefont {J.}~\bibnamefont
  {Gosner}}, \bibinfo {author} {\bibfnamefont {B.}~\bibnamefont {Kubala}}, \
  and\ \bibinfo {author} {\bibfnamefont {J.}~\bibnamefont {Ankerhold}},\ }\href
  {\doibase 10.1103/PhysRevB.101.054501} {\bibfield  {journal} {\bibinfo
  {journal} {Phys. Rev. B}\ }\textbf {\bibinfo {volume} {101}},\ \bibinfo
  {pages} {054501} (\bibinfo {year} {2020})}\BibitemShut {NoStop}%
\bibitem [{\citenamefont {Liang}\ \emph {et~al.}(2018)\citenamefont {Liang},
  \citenamefont {Marthaler},\ and\ \citenamefont {Guo}}]{Liang_2018}%
  \BibitemOpen
  \bibfield  {author} {\bibinfo {author} {\bibfnamefont {P.}~\bibnamefont
  {Liang}}, \bibinfo {author} {\bibfnamefont {M.}~\bibnamefont {Marthaler}}, \
  and\ \bibinfo {author} {\bibfnamefont {L.}~\bibnamefont {Guo}},\ }\href
  {\doibase 10.1088/1367-2630/aaa7c3} {\bibfield  {journal} {\bibinfo
  {journal} {New Journal of Physics}\ }\textbf {\bibinfo {volume} {20}},\
  \bibinfo {pages} {023043} (\bibinfo {year} {2018})}\BibitemShut {NoStop}%
\bibitem [{\citenamefont {Nathan}\ \emph {et~al.}(2020)\citenamefont {Nathan},
  \citenamefont {Refael}, \citenamefont {Rudner},\ and\ \citenamefont
  {Martin}}]{Nathan2020}%
  \BibitemOpen
  \bibfield  {author} {\bibinfo {author} {\bibfnamefont {F.}~\bibnamefont
  {Nathan}}, \bibinfo {author} {\bibfnamefont {G.}~\bibnamefont {Refael}},
  \bibinfo {author} {\bibfnamefont {M.~S.}\ \bibnamefont {Rudner}}, \ and\
  \bibinfo {author} {\bibfnamefont {I.}~\bibnamefont {Martin}},\ }\href
  {\doibase 10.1103/PhysRevResearch.2.043411} {\bibfield  {journal} {\bibinfo
  {journal} {Phys. Rev. Research}\ }\textbf {\bibinfo {volume} {2}},\ \bibinfo
  {pages} {043411} (\bibinfo {year} {2020})}\BibitemShut {NoStop}%
\bibitem [{\citenamefont {Kuramoto}(1975)}]{Kuramoto1975}%
  \BibitemOpen
  \bibfield  {author} {\bibinfo {author} {\bibfnamefont {Y.}~\bibnamefont
  {Kuramoto}},\ }\href {https://ci.nii.ac.jp/naid/10025994495/en/} {\bibfield
  {journal} {\bibinfo  {journal} {Lecture Notes in Physics}\ }\textbf {\bibinfo
  {volume} {30}},\ \bibinfo {pages} {420} (\bibinfo {year} {1975})}\BibitemShut
  {NoStop}%
\end{thebibliography}%


\onecolumngrid


\renewcommand\thefigure{S.\arabic{figure}}    
\setcounter{figure}{0}    

\renewcommand{\theequation}{S.\arabic{equation}}
\setcounter{equation}{0}

\section*{Supplemental Material}

\noindent
In this supplemental material to our article we present further details on (\textit{i.}) the theory of locking by direct ac-current injection into the cavity, deriving an effective Adler-equation similar to that presented in Sec.~\ref{sec:injection}; (\textit{ii.}) the derivation of the effective Kuramoto equations that supplements the results presented in Sec.~\ref{sec:synch}; and (\textit{iii.}) notes on the numerical calculations, relevant to all sections.



\section*{Locking from direct ac-current injection}
\begin{figure}[h]
\begin{center}
\includegraphics[width=0.5\columnwidth]{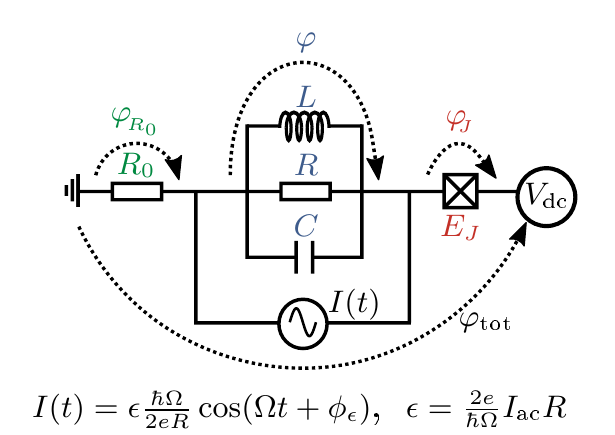}
\caption{
(Color online.) Sketch of a dc-voltage biased Josephson photonics circuit with an in-series resistance and a single resonance (with frequency $\omega_0=1/\sqrt{LC}$ and width $\gamma=1/RC$). A small ac-signal is injected by an external current source.
}
\label{fig:XCouplingCircuit}
\end{center}
\end{figure}

A similar locking scenario to the one described in the main text arises when directly injecting a locking signal into the cavity. Experimentally, this can be realized by feeding the cavity with an oscillating current through the transmission line. We now proceed to derive the locking equation for the circuit model of the Josephson photonics device shown in Fig. \ref{fig:XCouplingCircuit}.

Analogously to the main text, the classical equations of motions are the Kirchhoff equations
\begin{subequations}
\begin{align}
\dot\varphi_J = \frac{2e}{\hbar}V_\text{dc} - \frac{2e}{\hbar}I_c R_0 \sin(\varphi_J) - \dot{\varphi};\label{}\\
\ddot\varphi+\gamma\dot\varphi+\omega_0^2\varphi = \frac{2e I_c R}{\hbar\omega_0} \gamma\omega_0 \sin {\varphi_J} - \frac{2e}{\hbar} I(t).\label{}
\end{align}
\label{}
\end{subequations}
Here we assume the device is dc-biased by a fixed voltage $V_\text{dc}$. The injection locking signal is provided by the ac-current $I(t) = I_\text{ac} \cos(\Omega t+\phi_\epsilon)$ with amplitude assumed small, $\epsilon = (2e/\hbar)(I_\text{ac}R/\Omega)<1$. 
For the other quantities, we use the same parametrization as in Table \ref{table:1}. 
The locking equation of Adler-type is derived using the same time scale separation arguments as in the main text. We use a similar ansatz for the degrees of freedom $\varphi(t)$ and $\varphi_J(t)$ in terms of slowly-varying quantities 
\begin{subequations}
\begin{align}
\varphi_{J}(t) =&\ \omega_J t + \theta_J(t) + a_J \sin[\omega_J t+\phi_J(t)];\label{} \\
\varphi(t) =&\ a \sin[\omega_J t+\phi(t)]\; .\label{}
\end{align}
\end{subequations}
The unknown functions, $\theta_J(t),\,\phi_J(t)$ and $\phi(t)$, are slowly-varying in time, with $\dot\theta_J,\,\dot\phi_J,\,\dot\phi\simeq\nu\ll\omega_J$, while the `fast' 
frequencies are only slightly detuned from each other, $\omega_J\simeq\omega_\textrm{dc}\simeq\omega_0$.

Separating each equation of motion into slow and rapidly oscillating parts, we find the same equation for the slow component $\theta_J(t)$ as Eq.~\eqref{eq:slow1}, namely
\begin{align}
 \dot{\theta}_J = \omega_{\text{dc}} - \omega_J - v_{R_0} \frac{a_J}{2} \sin(\phi_J - \theta_J),
 \label{eq:Appslow}
\end{align}
The components oscillating with frequencies close to $\omega_J$ are modified in this biasing condition compared to the main text,
\begin{subequations}
\begin{align}
-\omega_J a_J \cos(\omega_Jt+\phi_J) =&\ v_{R_0}\sin(\omega_Jt+\theta_J) + a\omega_J\cos(\omega_Jt+\phi),\label{eq:Appfast1}\\
\tilde{I}_c\omega_0 \sin(\omega_Jt+\theta_J) - \epsilon \Omega \cos(\Omega t + \phi_{\epsilon})=&\ a\;\left\{ \frac{1}{\gamma}\left[\omega_0^2-\left(\omega_J+\dot\phi\right)^2\right]\sin(\omega_Jt+\phi)+
\omega_J\cos(\omega_Jt+\phi)\right\}.\label{eq:Appfast2}
\end{align} \label{eq:Appfast}
\end{subequations}
These equations are the counterparts of Eqs.~\eqref{eq:slow2} and \eqref{eq:slow3} in the main text. A transformation of eqs. (\ref{eq:Appfast}) to a frame rotating with $\omega_J$ yields
\begin{subequations}
\begin{align}
a_J e^{i\phi_J} =&\ i\frac{v_{R_0}}{\Omega}e^{i\theta_J} - a e^{i\phi} ,\label{eq:Approt1}\\
a e^{i\phi} =&\ \frac{1}{z\left(\nu_j\right)}\left(\tilde{I}_c e^{i\theta_J} - i \epsilon e^{i\phi_\epsilon}e^{i(\Omega-\omega_J)t}\right),\label{eq:Approt2}
\end{align} \label{eq:Approt}
\end{subequations}
where we have defined the complex dimensionless impedance $z(\nu_j)\equiv|z|e^{i\chi}=\left(2\nu_j+i\gamma\right)/\gamma$, in analogy to Sec.~\ref{sec:synch}, that here is always evaluated at the detuning $\nu_j\equiv \left(\omega_0-\omega_J-\dot\phi\right)$ between the effective Josephson frequency and the resonance. Outside the locking region $\nu_j$ is a function of the injected signal through $\omega_J(\epsilon)$ and $\dot\phi(\epsilon,t)$ and imprints its dependence onto the dimensionless impedance, both on its absolute value $|z|(\epsilon)$ and its phase $\chi(\epsilon)$. In the locking region $\dot\phi=0$ and $\omega_J=\Omega$, such that $\nu_j=\omega_0-\Omega$ becomes independent of $\epsilon$.

As in the main text, we find the locking equation by substituting Eq.~\eqref{eq:Approt2} into Eq.~\eqref{eq:Approt1} and taking the imaginary part to find an expression for $a_J \sin(\phi_J - \theta_J)$. The expression is then substituted into Eq.~\eqref{eq:Appslow}, yielding
\begin{align}
\dot\theta_J = \left(\omega_\text{dc} - \omega_J\right) - \frac{v_{R_0}}{2}
\left[\epsilon\: \frac{1}{|z|}\sin\left[(\Omega-\omega_J)t-\theta_J+\phi_\epsilon-\chi+\frac{\pi}{2}\right]  + \frac{v_{R_0}}{\Omega} + \frac{\tilde{I}_c}{|z|}\sin(\chi) \right].\label{eq:AppAdler1}
\end{align}
which is analogous to Eq.~\eqref{eq:Adler1} in the main text.

Defining the Adler phase similarly to Eq.~\eqref{eq:AdlerPsi},
\begin{align}
\psi(t) = \omega_Jt+\theta_J - \Omega t-\phi_\epsilon-\frac{\pi}{2},
\end{align}
we obtain the locking equation for a direct ac-current injection. Here too the locking equation has the form of a generalized Adler equation,
\begin{align}
\dot\psi =&\ \nu(\epsilon,\Omega) - \nu_c(\epsilon,\Omega)\;\frac{ \epsilon}{2} \sin\left[\psi + \chi(\epsilon,\Omega)\right].\label{eq:AppAdlerfin}
\end{align}
The parameters are given by,
\begin{subequations}
\begin{align}
\nu(\epsilon,\Omega) = &\ \left(\omega_\text{dc} -  \frac{v_{R_0}}{\Omega}\: \frac{v_{R_0}}{2} - \frac{\tilde{I}_c}{|z|(\epsilon,\Omega)}\;\frac{v_{R_0}}{2} \right)- \Omega.\label{eq:AppAdnu} \\
\nu_c(\epsilon,\Omega) =&\ \frac{v_{R_0}}{|z|(\epsilon,\Omega)}, \quad |z|(\epsilon,\Omega) = \frac{1}{\gamma}\sqrt{4\left[\omega_0-\omega_J(\epsilon,\Omega)-\dot\phi(\epsilon,\Omega)\right]^2+\gamma^2}.\label{eq:AppAdnuc}\\
e^{i\chi(\epsilon,\Omega)} =&\ \frac{z(\epsilon,\Omega)}{|z|(\epsilon,\Omega)} = \frac{2\left[\omega_0-\omega_J(\epsilon,\Omega)-\dot\phi(\epsilon,\Omega)\right]+i\gamma}{\sqrt{4\left[\omega_0-\omega_J(\epsilon,\Omega)-\dot\phi(\epsilon,\Omega)\right]^2+\gamma^2}}
\end{align}
\end{subequations}
Compared to Eq.~\eqref{eq:Adfin}, here not only the effective detuning $\nu(\epsilon,\Omega)$, but also the effective width of the locking region $\nu_c(\epsilon,\Omega)$ and the effective locked phase $\chi(\epsilon,\Omega)$ acquire dependence on the injection parameters $\Omega$ and $\epsilon$ through the dimensionless impedance $z(\nu_j)$.

\section*{Derivation of the effective Kuramoto-type equations for synchronization}
The derivation of the Kuramoto-type equations for synchronization starts from the full circuit equations, Eq.~\eqref{eq:EOM-synch},  
and the ansatz for the dominant oscillations, Eqs.~\eqref{eq:ansatz_sync_JJ} and \eqref{eq:ansatz_sync_cav} consistent with the limit of weak Josephson coupling $\tilde I_{c}^{(\sigma)}\ll 1$ and small low-frequency impedance $R_0^{(\sigma)}$, i.e. $v_{R_0}^{(\sigma)}\ll \omega_0^{(\sigma)}$. 
We further assume time scale separation 
$\dot\theta_{J}^{(\sigma)},\dot\phi_{J}^{(\sigma)},\dot\phi^{(\sigma)}\ll\omega_\text{dc}^{(\sigma)},\omega_{0}^{(\sigma)},\omega_{J}^{(\sigma)}$, 
as well as $(\omega_\text{dc}^{(\sigma)}-\omega_{0}^{(\sigma)}),\left(\omega^{(2)}_\text{dc}-\omega^{(1)}_\text{dc}\right)\ll\omega_\text{dc}^{(\sigma)},\omega_{0}^{(\sigma)},\omega_{J}^{(\sigma)}$.
The slowly-varying functions obey the following system of coupled equations
\begin{subequations}
\begin{align}
\dot\theta_{J}^{(1)} =&\  \left(\omega_\text{dc}^{(1)}-\omega_{J}^{(1)}\right) - v_{R_0}^{(1)}\frac{a^{(1)}}{2}\sin\left(\phi_{J}^{(1)}-\theta_{J}^{(1)}\right), 
\label{eq:c1}\\
\dot\theta_{J}^{(2)} =&\  \left(\omega_\text{dc}^{(2)}-\omega_{J}^{(2)}\right) - v_{R_0}^{(2)}\frac{a^{(2)}}{2}\sin\left(\phi_{J}^{(2)}-\theta_{J}^{(2)}\right), 
\label{eq:c2}\\
z^{(1)} b^{(1)} e^{i\phi^{(1)}} =&\ \tilde I_{c}^{(1)}e^{i\theta_{J}^{(1)}}-\epsilon^{(1)}\frac{\omega_{J}^{(2)}}{\gamma^{(1)}}b^{(2)}e^{i\phi^{(2)}}e^{i\nu_Jt},
\label{eq:c3}\\
z^{(2)} b^{(2)} e^{i\phi^{(2)}} =&\  \tilde I_{c}^{(2)}e^{i\theta_{J}^{(2)}}-\epsilon^{(2)}\frac{\omega_{J}^{(1)}}{\gamma^{(2)}}b^{(1)}e^{i\phi^{(1)}}e^{-i\nu_Jt},
\label{eq:c4}\\
a^{(1)}e^{i\phi_J^{(1)}} =&\  i\frac{v_{R_0}^{(1)}}{\omega_0^{(1)}}e^{i\theta_{J}^{(1)}}- b^{(1)} e^{i\phi^{(1)}}, 
\label{eq:c5}\\
a^{(2)}e^{i\phi_J^{(2)}} =&\  i\frac{v_{R_0}^{(2)}}{\omega_0^{(2)}}e^{i\theta_{J}^{(2)}}- b^{(2)} e^{i\phi^{(2)}}. 
\label{eq:c6}
\end{align}
\end{subequations}
where $\nu_J=\left(\omega_{J}^{(2)}-\omega_{J}^{(1)}\right)$ is the detuning between the Josephson oscillations of the two devices. 
We have also introduced the complex dimensionless impedance 
$z^{(\sigma)}=\left(2\nu^{(\sigma)}+i\gamma^{(\sigma)}\right)/\gamma^{(\sigma)}$, that is obtained from the Fourier transform $\tilde{Z}^{(\sigma)}(\omega)$ of the response $Z^{(\sigma)}$. 
The impedance $z^{(\sigma)} = \tilde{Z}^{(\sigma)}(\omega= \omega_0^{(\sigma)}-\nu^{(\sigma)})$ is evaluated at the detuning $\nu^{(\sigma)}$ given by
$\nu^{(\sigma)}\equiv \left(\omega_{0}^{(\sigma)} -\omega_{J}^{(\sigma)}-\dot\phi^{(\sigma)}\right)$.

Eqs.~\eqref{eq:c3} and~\eqref{eq:c4} can be written as a
matrix equation for the vector $v\equiv \left[b^{(1)}e^{i\phi^{(1)}},b^{(2)}e^{i\phi^{(2)}}\right]^T$ containing the cavity oscillation amplitudes,
\begin{align}
M v =&\ v_J,\quad \textrm{with}\quad 
M\equiv 
\begin{bmatrix}
z^{(1)} & \epsilon^{(1)}\frac{\omega_{J}^{(2)}}{\gamma^{(1)}}e^{i\nu_J t}\\
\epsilon^{(2)}\frac{\omega_{J}^{(1)}}{\gamma^{(2)}}e^{-i\nu_J t} & z^{(2)}
\end{bmatrix}, 
\quad \textrm{and}\quad 
v_J \equiv 
\left[\tilde I_{c}^{(1)}e^{i\theta_J^{(1)}},\tilde I_{c}^{(2)}e^{i\theta_J^{(2)}}\right]^T.\notag
\end{align}
The matrix $M$ can be inverted analytically to obtain the following expressions for the cavity amplitudes
\begin{subequations}
\begin{align}
b^{(1)} e^{i\phi^{(1)}} =&\ \frac{\tilde I_{c}^{(1)}e^{i\theta_{J}^{(1)}} z^{(2)}-\epsilon^{(1)} \left(\omega_{J}^{(2)}/\gamma^{(1)}\right)\tilde I_{c}^{(2)}e^{i\theta_J^{(2)}}e^{i\nu_J t}}{z^{(1)}z^{(2)}-\epsilon^{(1)}\epsilon^{(2)}\left(\omega_{J}^{(1)}/\gamma^{(1)}\right)\left(\omega_{J}^{(2)}/\gamma^{(2)}\right)},\\
b^{(2)} e^{i\phi^{(2)}} =&\ \frac{\tilde I_{c}^{(2)}e^{i\theta_{J}^{(2)}} z^{(1)}-\epsilon^{(2)} \left(\omega_{J}^{(1)}/\gamma^{(2)}\right)\tilde I_{c}^{(1)}e^{i\theta_J^{(1)}}e^{-i\nu_J t}}{z^{(1)}z^{(2)}-\epsilon^{(1)}\epsilon^{(2)}\left(\omega_{J}^{(1)}/\gamma^{(1)}\right)\left(\omega_{J}^{(2)}/\gamma^{(2)}\right)}.
\end{align}
\end{subequations}
The above expressions for the amplitudes of the two cavity oscillations can be inserted into the original 
set of coupled equations, specifically Eqns.~\eqref{eq:c5} and \eqref{eq:c6},
to obtain expressions for the relative phases $(\theta_J^{(\sigma)}-\phi_J^{(\sigma)})$, that can then be inserted 
back into Eqns.~\eqref{eq:c1} and \eqref{eq:c2}, in analogy to the procedure used in Sec.~\ref{sec:injection}. 
We arrive at the following two equations
for the slow components of the Josephson phases,
\begin{subequations}
\begin{align}
\dot\theta_J^{(1)} =&\  \left(\omega_\text{dc}^{(1)}-\omega_{J}^{(1)}\right) - v_{R_0}^{(1)}\frac{1}{2}\text{Im}\left\{i\frac{v_{R_0}^{(1)}}{\omega_0^{(1)}}- \frac{\tilde I_{c}^{(1)} z^{(2)}-\epsilon^{(1)} \left(\omega_{J}^{(2)}/\gamma^{(1)}\right)\tilde I_{c}^{(2)}e^{i\nu_J t}e^{i\left(\theta_J^{(2)}-\theta_J^{(1)}\right)}}{z^{(1)}z^{(2)}-\epsilon^{(1)}\epsilon^{(2)}\left(\omega_{J}^{(1)}/\gamma^{(1)}\right)\left(\omega_{J}^{(2)}/\gamma^{(2)}\right)}\right\}, \label{eq:J2ab1}\\
\dot\theta_J^{(2)} =&\  \left(\omega_\text{dc}^{(2)}-\omega_{J}^{(2)}\right) - v_{R_0}^{(2)}\omega_{0}^{(2)}\frac{1}{2}\text{Im}\left\{i\frac{v_{R_0}^{(2)}}{\omega_0^{(2)}}- \frac{\tilde I_{c}^{(2)} z^{(1)}-\epsilon^{(2)} \left(\omega_{J}^{(1)}/\gamma^{(2)}\right)\tilde I_{c}^{(1)}e^{-i\nu_J t}e^{i\left(\theta_J^{(1)}-\theta_J^{(2)}\right)}}{z^{(1)}z^{(2)}-\epsilon^{(1)}\epsilon^{(2)}\left(\omega_{J}^{(1)}/\gamma^{(1)}\right)\left(\omega_{J}^{(2)}/\gamma^{(2)}\right)}\right\}. \label{eq:J2ab2}
\end{align}
\end{subequations}
The above equations give the non-linear evolution of the slow phases $\theta_{J}^{(\sigma)}$ as a function of the couplings $\epsilon^{(\sigma)}$.
These equations reduce to the Kuramoto model in the limit
$\epsilon^{(1)}\left(\omega_{J}^{(2)}/\gamma_{(1)}\right) \ll 1$ and
$\epsilon^{(2)} \left(\omega_{J}^{(1)}/\gamma_{(2)}\right) \ll 1$, 
where after linearizing with respect to the couplings, the equations become
\begin{subequations}
\begin{align}
\dot\theta_J^{(1)} =&\  \tilde\nu^{(1)}+\epsilon^{(1)}\ \frac{1}{2}v_{R_0}^{(1)}\frac{\omega_{J}^{(1)}}{\gamma^{(2)}}
\frac{\tilde I_{c}^{(2)}}{|z_1||z_2|} \sin\left[\nu_Jt+\theta_J^{(2)}-\theta_J^{(1)}-\chi_1-\chi_2\right], \label{eq:lin1}\\
\dot\theta_J^{(2)} =&\  \tilde\nu^{(2)} -\epsilon^{(2)}\ \frac{1}{2}v_{R_0}^{(2)}\frac{\omega_{J}^{(2)}}{\gamma^{(1)}}
\frac{\tilde I_{c}^{(1)}}{|z_1||z_2|}\sin\left[\nu_Jt+\theta_J^{(2)}-\theta_J^{(1)}+\chi_1+\chi_2\right]. \label{eq:lin2}
\end{align}
\end{subequations}
with $\tilde\nu^{(\sigma)}\equiv \omega_J^{(\sigma)}(\epsilon=0)-\omega_J^{(\sigma)}$ and $z^{(\sigma)}\equiv |z^{(\sigma)}| e^{\chi^{(\sigma)}}$.
These equations are equivalent to Eq.~\eqref{eqMT:lin} of the main text.

\section*{Some notes on numerical implementation}

For numerical results the coupled equations of motions, Eq.~\eqref{eq:eom}, were solved using a real-valued variable-coefficient ordinary differential equation (VODE) solver with a BDF method implemented in the Python library SciPy.  
Typically solutions were calculated for  
time intervals of $0\leq t\omega_0 \leq 10^5$.  
Spectra were calculated using standard FFT routines from SciPy with a frequency resolution of $\delta \omega /  \omega_0 \approx 8 \cdot 10^{-5}$ given by a time interval $2.5 \cdot 10^4 < t\omega_0 \leq 10^5$ after reaching the steady state. To regularize the spectra we used a Kaiser-Bessel window with a shape parametrized by $\alpha=3$.
The Josephson frequency was numerically computed by a time average of the solution for $\dot{\varphi}_J$ in a time interval $7.5 \cdot 10^4 \leq t\omega_0 \leq 10^5$.

Simulations including noise use a lower-order Euler-Maruyama algorithm to solve the full equations of motions with an included auxiliary equation creating colored noise by an Ornstein-Uhlenbeck process. Wiener increments are drawn from a Gaussian distribution with random seed by a NumPy random number generator. Here were used a rectangular window function to calculate spectra.

Phase space distributions were calculated in a rotating frame with $200$ bins in both coordinate directions. We used a larger steady state interval $2.5 \cdot 10^5 \leq t\omega_0 \leq 10^6$ with $7.5 \cdot 10^7$ time steps ($2 \cdot 10^5 \leq t\omega_0 \leq 7.2 \cdot 10^5$ with $14 \cdot 10^7$ time steps for Fig.~\ref{fig:parametric_fast_lock_sig}(c) respectively).

With these parameters plots can be easily created without extensive optimization to reduce numerical costs. Single runs on standard PCs, or two-parameters sweeps and multi-runs for noise averaging on a Baden-W\"urttemberg Cluster JUSTUS2 require typical runtimes ranging from few minutes to a few days on $\sim 100$ cores.

\end{document}